\documentclass{natureprintstyle}
\usepackage[utf8]{inputenc}
\usepackage[super,comma,sort&compress]{natbib}
\usepackage{graphicx}
\usepackage{amssymb}
\usepackage{lineno}
\usepackage{xpatch}
\usepackage[T1]{fontenc}  % needed for the 'bar' bit below
\usepackage[labelfont=bf]{caption}
\DeclareCaptionLabelSeparator{bar}{ | }
\let\thebibliographytwo\thebibliography%
\usepackage{etoolbox}
\makeatletter
\apptocmd{\thebibliographytwo}{\global\c@NAT@ctr 34\relax}{}{}
\makeatother

\makeatletter
\renewcommand{\fnum@figure}{Fig. \thefigure}
\makeatother

\newcommand\araa{Annu. Rev. Astron. Astrophys.}%  % Annual Review of Astron and Astrophys 
\newcommand\apj{Astrophys. J.}%    % Astrophysical Journal 
\newcommand\apjl{Astrophys. J. Lett.}%     % Astrophysical Journal, Letters 
\newcommand\apjs{Astrophys. J. Suppl. Ser.}%    % Astrophysical Journal, Supplement 
\newcommand\aap{Astron. Astrophys.}%     % Astronomy and Astrophysics 
\newcommand\aapr{Astron. Astrophys. Rev.}%  % Astronomy and Astrophysics Reviews 
\newcommand\mnras{Mon. Notices R. Astron. Soc.}%   % Monthly Notices of the RAS 
\newcommand\nat{Nature}
\newcommand\aj{Astron. J.}

\makeatletter
\let\saved@includegraphics\includegraphics
\AtBeginDocument{\let\includegraphics\saved@includegraphics}
\renewenvironment*{figure}{\@float{figure}}{\end@float}
\makeatother

\title{A population of ultraviolet-dim protoclusters detected in absorption}

\author{Andrew B. Newman${}^1$, Gwen C. Rudie${}^1$, Guillermo A. Blanc${}^{1,2}$, Mahdi Qezlou${}^{1,3}$, Simeon Bird${}^3$, Daniel D. Kelson${}^1$, Victoria P\'{e}rez${}^2$, Enrico Congiu${}^2$, Brian C. Lemaux${}^{4,5}$, Alan Dressler${}^1$, and John S. Mulchaey${}^1$ \\[1ex]
${}^{1}$ Observatories of the Carnegie Institution for Science, 813 Santa Barbara Street, Pasadena, California 91101, USA\\
${}^2$ Departamento de Astronom\'{i}a, Universidad de Chile, Camino del Observatorio 1515, Las Condes, Santiago, Chile\\
${}^3$ Department of Physics and Astronomy, University of California, Riverside, 900 University Avenue, Riverside, California 92521, USA\\
${}^4$ Department of Physics and Astronomy, University of California, Davis, One Shields Avenue, Davis, California 95616, USA\\
${}^5$ Gemini Observatory, NSF’s NOIRLab, 670 N. A’ohoku Place, Hilo, Hawai’i, 96720, USA}
\begin{document}

\maketitle

\begin{abstract}
Galaxy protoclusters, which will eventually grow into the massive clusters we see in the local universe, are usually traced by locating overdensities of galaxies.\citep{Overzier16} Large spectroscopic surveys of distant galaxies now exist, but their sensitivity depends mainly on a galaxy’s star formation activity and dust content rather than its mass. Tracers of massive protoclusters that do not rely on their galaxy constituents are therefore needed. Here we report observations of Lyman-$\alpha$ absorption in the spectra of a dense grid of background galaxies,\citep{Lee14,Newman20} which we use to locate a substantial number of candidate protoclusters at redshifts 2.2-2.8 via their intergalactic gas. We find that the structures producing the most absorption, most of which were previously unknown, contain surprisingly few galaxies compared to the dark matter content of their analogs in cosmological simulations.\citep{Klypin16,Nelson19} Nearly all are expected to be protoclusters, and we infer that half of their expected galaxy members are missing from our survey because they are unusually dim at rest-frame ultraviolet wavelengths. We attribute this to an unexpectedly strong and early influence of the protocluster environment\citep{Contini16,Muldrew18} on the evolution of these galaxies that reduced their star formation or increased their dust content.
%\blfootnote{\noindent\begin{affiliations}
%\hspace{-0.7cm} \item Observatories of the Carnegie Institution for Science, 813 Santa Barbara Street, Pasadena, California 91101, USA
%\item Departamento de Astronom\'{i}a, Universidad de Chile, Camino del Observatorio 1515, Las Condes, Santiago, Chile
%\item Department of Physics \& Astronomy, University of California, Riverside, 900 University Avenue, Riverside, California 92521, USA
%\item Department of Physics and Astronomy, University of California, Davis, One Shields Avenue, Davis, California 95616, USA
%\item Gemini Observatory, NSF’s NOIRLab, 670 N. A’ohoku Place, Hilo, Hawai’i, 96720, USA
%\end{affiliations}}
\end{abstract}

We have mapped Lyman-$\alpha$ (Ly$\alpha$) absorption over 1.5~deg${}^2$ and the redshift interval $z=2.2$-2.8 via the Lyman-$\alpha$ Tomography IMACS Survey (LATIS),\citep{Newman20} conducted at the Magellan Baade telescope at Las Campanas Observatory. The survey comprises deep spectra of a dense sample of Lyman-break galaxies and quasars. We reconstructed the three-dimensional (3D) distribution of Ly$\alpha$ absorption imprinted in these spectra by neutral hydrogen in the intergalactic medium (IGM), a technique known as Ly$\alpha$ or IGM tomography.\citep{Pichon01,Lee14} This absorption traces the density field on large scales, and the resulting maps are expected to be particularly effective at locating protoclusters.\citep{Stark15} LATIS spans three extragalactic deep fields and exceeds the volume of the pioneering Ly$\alpha$ tomography survey CLAMATO\citep{Lee18} by a factor of 10, making it uniquely suited to the discovery of rare structures like protoclusters. The key survey products are 3D maps of the Ly$\alpha$ flux contrast $\delta_F = F / \langle F \rangle - 1$ smoothed to a resolution of 4 $h^{-1}$ comoving Mpc (cMpc), where $F$ is the continuum-normalized Ly$\alpha$ flux (Fig.~1), along with 2,241 secure redshifts of Lyman-break galaxies and quasars within the same volumes. These galaxies are selected based on either their photometric redshifts or optical colors\citep{Newman20} and $r$-band magnitudes $r < 24.8$. Their median rest-frame ultraviolet (UV) luminosity density is $\approx L^{*}$. We refer to these as UV-selected galaxies, because the LATIS selection is typical of techniques based on the rest-UV (observed-frame optical) continuum emission, which traces unobscured star formation.

\begin{figure*}[ht]
    \centering
    \includegraphics[width=5.25in]{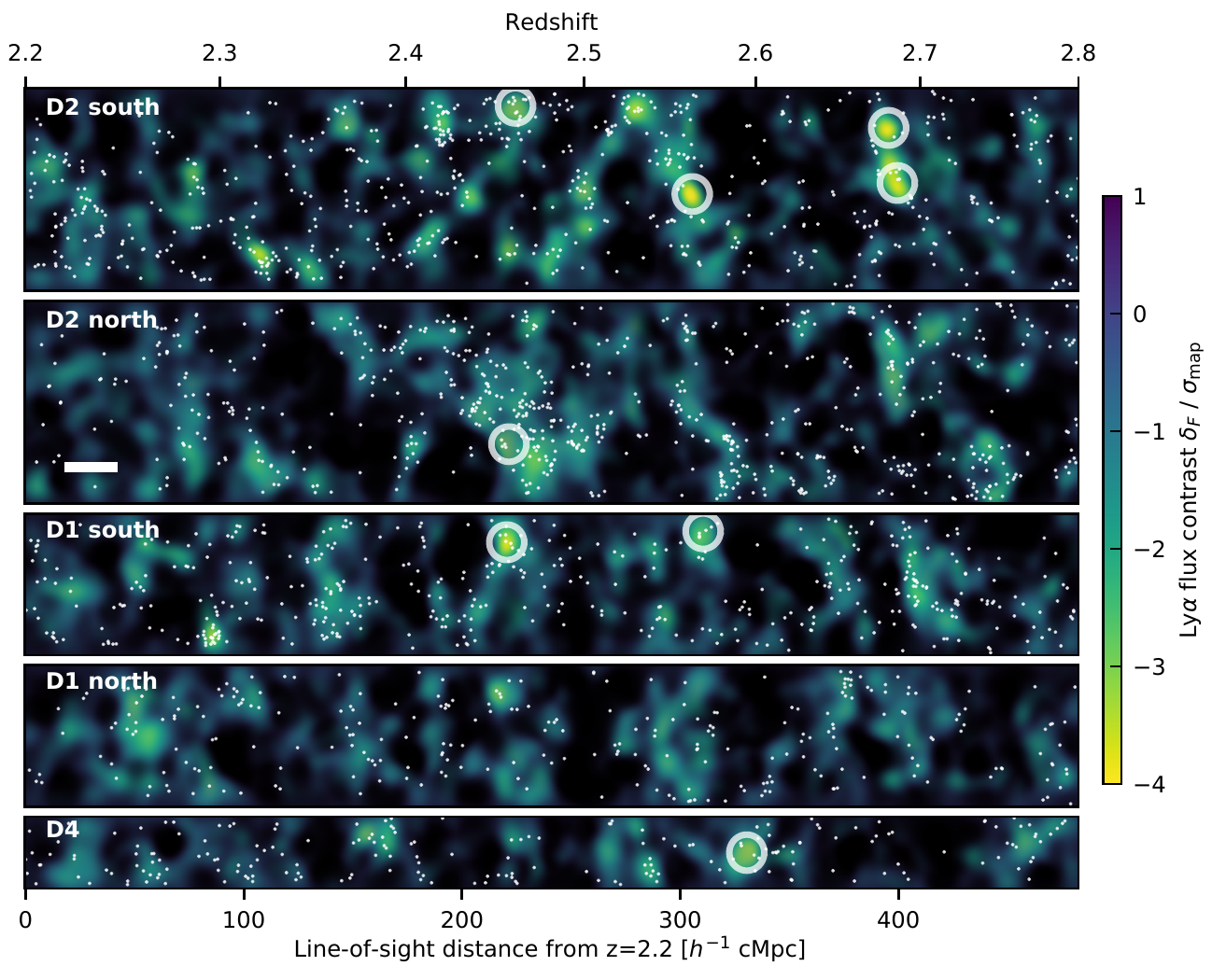}
    \caption{{\bf Maps of the intergalactic medium.} Each panel shows a three-dimensional volume rendering viewed along the declination axis. Structures are rendered semi-transparent with color encoding the Ly$\alpha$ flux contrast. Points show the positions of LATIS galaxies. The maps are smoothed by a Gaussian kernel with $\sigma = 4$~$h^{-1}$~cMpc to maximize sensitivity to protoclusters.\citep{Stark15} The three survey fields are shown, with D1 and D2 (COSMOS) split into their northern and southern halves. The scale bar is 20 $h^{-1}$ cMpc. The eight strongest absorption peaks are encircled.}
    \label{fig:map}
\end{figure*}

We identified a large sample of matter overdensities in the IGM maps and examined their galaxy contents. Following previous work,\citep{Stark15,Lee16} we identified these overdensities as local minima in the smoothed Ly$\alpha$ flux maps. We found 149 such minima having $\delta_F / \sigma_{\rm map} < -2$, which we refer to as absorption peaks. (The map standard deviation $\sigma_{\rm map}$ is dominated by IGM structure rather than observational noise, which is typically $1.8\times$ smaller.\citep{Newman20}) We interpret the nature of these absorption peaks using a suite of mock LATIS surveys performed in cosmological simulations. Each mock survey realization produces tomographic maps constructed from mock spectra that follow the exact spatial distribution of the LATIS sightlines and their noise properties. The mock maps accurately match the statistical properties of the observed ones. On average, stronger absorption peaks are associated with higher matter overdensities and with the progenitors of more massive $z=0$ halos in the mock surveys (Methods), as previously reported.\citep{Lee16} Based on the statistical distribution in the mock surveys, among the 149 observed absorption peaks with $\delta_F / \sigma_{\rm map} < -2$ we expect that 76 are progenitors of a galaxy cluster ($M_{z=0} / {\rm M}_{\odot} > 10^{14}$) and 43 of a massive group ($10^{13.5} < M_{z=0} / {\rm M}_{\odot} < 10^{14}$), on average. Among the 29 absorption peaks with $\delta_F/\sigma_{\rm map} < -3$, we expect 23 protoclusters and 4 protogroups.

At the location of each absorption peak, we measured the overdensity $\delta_{\rm gal}$ of LATIS galaxies within 8 $h^{-1}$ cMpc. The connection between the galaxy overdensity and Ly$\alpha$ absorption is shown in Fig.~2. As the absorption increases ($\delta_F$ more negative), the galaxy overdensity initially increases as expected. However, the relationship is not monotonic: the strongest absorption peaks, which should correspond to the highest matter overdensities, show a statistically significant but surprisingly small galaxy overdensity ($\delta_{\rm gal} \approx 1$), on average.

To further examine this unanticipated result, we calculated the overdensity of dark matter halos $\delta_{\rm halo}$ in the mock surveys at the location of each simulated absorption peak.  We considered halos with masses $M_{\rm vir} > 10^{11.8} M_{\odot}$, which match the autocorrelation function of the LATIS galaxies (Methods). These halos were randomly sampled to match the space density of the LATIS galaxies. 
The expected trend between Ly$\alpha$ absorption and galaxy overdensity, assuming that $\delta_{\rm gal} = \delta_{\rm halo}$, is illustrated by the green bands in Fig. 2. The observed trend matches this expectation closely for the majority of absorption peaks having $\delta_F / \sigma_{\rm map} \gtrsim -3.5$. But this agreement breaks down in the strongest absorption peaks, where the observed $\delta_{\rm gal}$ becomes smaller than $\delta_{\rm halo}$ in all but $p=0.005$\% of the mock surveys; such a significant difference occurs at any value of $\delta_F$ in only $p=0.1$\% of mock surveys. The observed trend in Fig.~2 is thus unlikely if UV-selected galaxies trace halos independently of the large-scale environment.

The unexpectedly low $\delta_{\rm gal}$ is most evident in the eight strongest absorption peaks in LATIS with $\delta_F / \sigma_{\rm map} < -3.8$. According to their IGM signature, 94\% of these are protoclusters (Methods). Yet only the two with the highest $\delta_{\rm gal}$ are definitely associated with previously known protoclusters.\citep{Chiang14,Chiang15,Diener15,Casey15,Cucciati18} The remaining six systems have not been detected as galaxy overdensities, even though they lie in some of the most thoroughly observed extragalactic deep fields. Importantly, we find that these strong absorption peaks are not themselves unusual, only their galaxy contents. They occur in the LATIS maps with the expected frequency, and they arise from widespread absorption across most of the 10 sightlines that typically probe each structure within 4 $h^{-1}$ cMpc. This and other arguments (Methods) make it clear that contamination by high-column density absorbers associated with galaxies, rather than Mpc-scale overdensities, is not significant. Likewise, locally enhanced ionization (e.g., from quasar radiation) could reduce Ly$\alpha$ absorption, but it cannot account for a low galaxy abundance within strong absorption peaks.

We conclude that most of the strongest Ly$\alpha$ absorption peaks are genuine protoclusters containing an enhanced population of dark matter halos that do not host UV-selected galaxies. By examining the radial profile of $\delta_{\rm gal}$ (Fig.~3), we find that the dearth of such galaxies is most prominent toward the centers of the strongest absorption peaks, where the fraction of halos hosting UV-selected galaxies reaches about 50\% of its value in other environments. The observed-frame optical and near-infrared colors and luminosities of those LATIS galaxies that are found within the strong absorption peaks are not significantly different from the overall population. This is consistent with some other studies showing little or no difference between Lyman-break galaxies in protoclusters and the field.\citep{Cucciati14,Lemaux14} At lower redshifts $z \lesssim 1$, the effect of the environment is mainly to modulate the mix of galaxy populations (e.g., star-forming and quiescent galaxies) rather than the properties of galaxies with one class.\citep{Blanton09,Muzzin12} Analogously, one possibility is that the ``missing'' galaxies in the strongest LATIS absorption peaks are UV-dim because they are dusty or quiescent.

\begin{figure}
    \centering
    \includegraphics[width=3.5in]{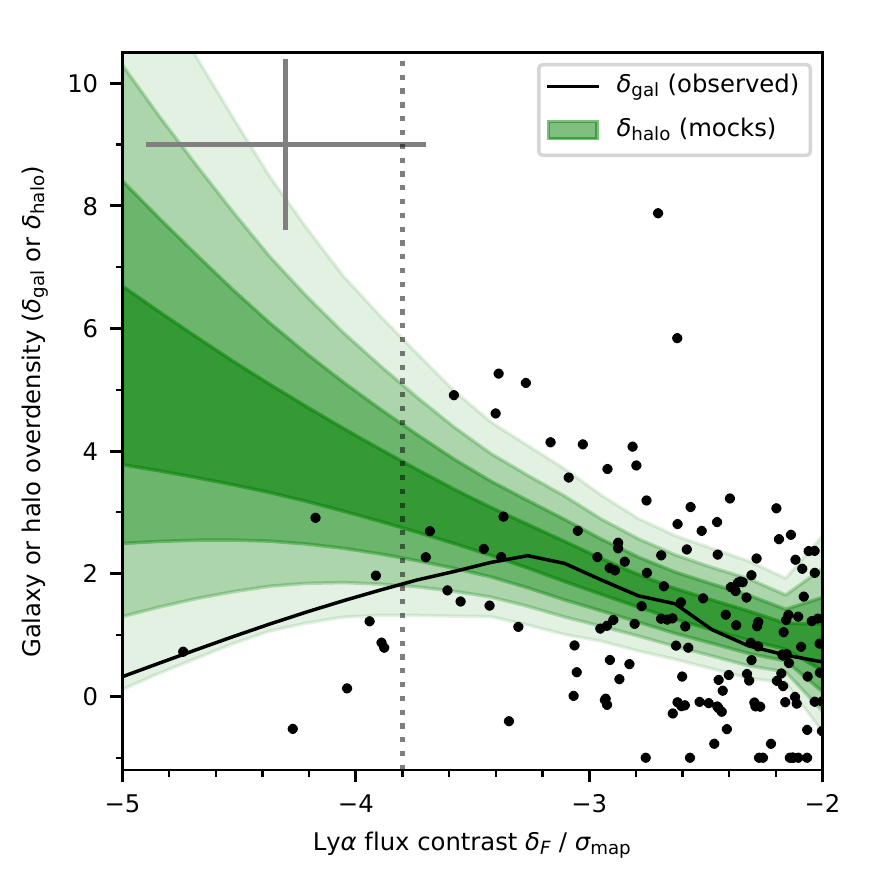}
    \caption{{\bf The galaxy overdensity $\delta_{\rm gal}$ at the location of Ly$\alpha$ absorption peaks.}  Black points indicate individual observed absorption peaks, with a representative 1 s.e.~error bar shown. More negative $\delta_F$ indicates more absorption. The black trendline\citep{Cappellari13} is compared to the distribution of trendlines (green bands indicate percentiles equivalent to 1, 2, 3 and 4 s.d.~in a normal distribution) in mock surveys that mimic LATIS within a cosmological simulation. In the simulations, we consider the overdensity of dark matter halos $\delta_{\rm halo}$ with $M_{\rm vir} > 10^{11.8} M_{\odot}$. Overdensities are measured within a cylinder that has a transverse radius and half-depth equal to 8 $h^{-1}$ cMpc. In the strongest absorption peaks (left of vertical line), $\delta_{\rm gal}$ is much lower than expected.}
    \label{fig:deltas}
\end{figure}

Indeed, previous studies have argued that dusty star-forming galaxies are remarkably abundant in many protoclusters\citep{Casey16} and that optical- and UV-dark galaxies are the dominant type at high stellar masses $M_* \gtrsim 10^{10.5} {\rm M}_{\odot}$ and redshifts $z > 3$.\citep{Wang19,Shu22} There is also evidence for a boosted fraction of quenched\citep{Wang16,McConachie21} or quenching\citep{Zavala19} massive galaxies in a few protoclusters. However, these relatively massive and rare galaxies usually occur alongside a substantial overdensity of UV-selected galaxies. A few studies have identified intriguing cases where UV-selected galaxies appear less numerous than expected based on other tracers\citep{Chapman09,Hung16,Shi19}, possibly analogous to the UV-dim LATIS structures. The key advance of our study is to detect an ensemble of protoclusters independent of their galaxy content, statistically compare their observed members to the underlying halo population, and thereby infer the presence of galaxies not yet detected in spectroscopic surveys.

The mass distribution of halos hosting LATIS galaxies is expected to be only weakly sensitive to the environment (Methods). This suggests that the distinct properties of galaxies in the strong absorption peaks have a more subtle environmental origin. Quiescent galaxies are expected in semianalytic models to be more common in protoclusters, preferentially at low stellar masses, due to the higher fraction of satellite galaxies and the earlier formation of halos. But this enhancement is predicted to be modest, at most $\sim$20\%.\citep{Contini16,Muldrew18} Our mock surveys based on the IllustrisTNG simulation (Methods) likewise show a very modest increment in the quiescent fraction (a few percentage points) within the strongest Ly$\alpha$ absorption peaks. Observations at lower redshifts suggest that the quiescent fraction of cluster galaxies reaches 50\% by $z \sim 1.5$\citep{Nantais17} with substantial cluster-to-cluster scatter. Thus, if quiescent galaxies account for the missing population in the strongest LATIS absorption peaks, they must be more abundant than theoretically anticipated, and they likely quenched earlier than most cluster galaxies. That would be consistent with faster galaxy evolution occurring in the most evolved protoclusters, since the strongest Ly$\alpha$ absorption peaks are expected to trace the densest regions on 4 $h^{-1}$ cMpc scales.

It is also possible that the galaxies missing from the strongest LATIS absorption peaks are moderately UV-dim, rather than being extremely dusty or quiescent. Based on photometric redshift catalogs,\citep{Weaver22} we estimate that two-thirds of galaxies selected to match the stellar mass distribution of the LATIS sample are fainter than the rest-UV flux limit of our survey and similar ones. The spectral energy distributions of such galaxies indicate that most are only mildly less star-forming or more dusty than their UV-brighter counterparts. An overabundance of such galaxies might account for the low overdensities we observe within the strongest LATIS absorption peaks.

Further observations are needed to search for the implied missing galaxy population. 
This search is of great interest, but it will be difficult. Although photometric redshifts can, in principle, expand the range of galaxies beyond that accessible with rest-UV spectroscopy, our simulations show that they lack the precision needed to test the trend in Fig.~2 (Methods). Spectroscopic redshifts are thus necessary, but they are very sparse for UV-faint galaxies.
Moreover, in order to substantially affect $\delta_{\rm gal}$, a ``missing'' population must be abundant and thus cannot have stellar masses far exceeding the LATIS galaxies' median $M_* \approx 10^{9.8} {\rm M}_{\odot}$.\citep{Weaver22} A spectroscopic redshift of a quiescent galaxy in this mass and redshift regime has never been obtained, demonstrating the challenge of assembling a complete inventory. Observations with the Atacama Large Millimeter Array (ALMA) will soon test whether sub-mm-selected galaxies reside in two of the strongest LATIS absorption peaks in COSMOS. Since our findings pertain to the average properties of absorption peaks, which individually have substantial uncertainties in $\delta_F$ and $\delta_{\rm gal}$ and are expected to span a range of overdensities and descendant masses (Methods), definitive follow-up observations must encompass an ensemble of such systems.

\begin{figure}
    \centering
    \includegraphics[width=88mm]{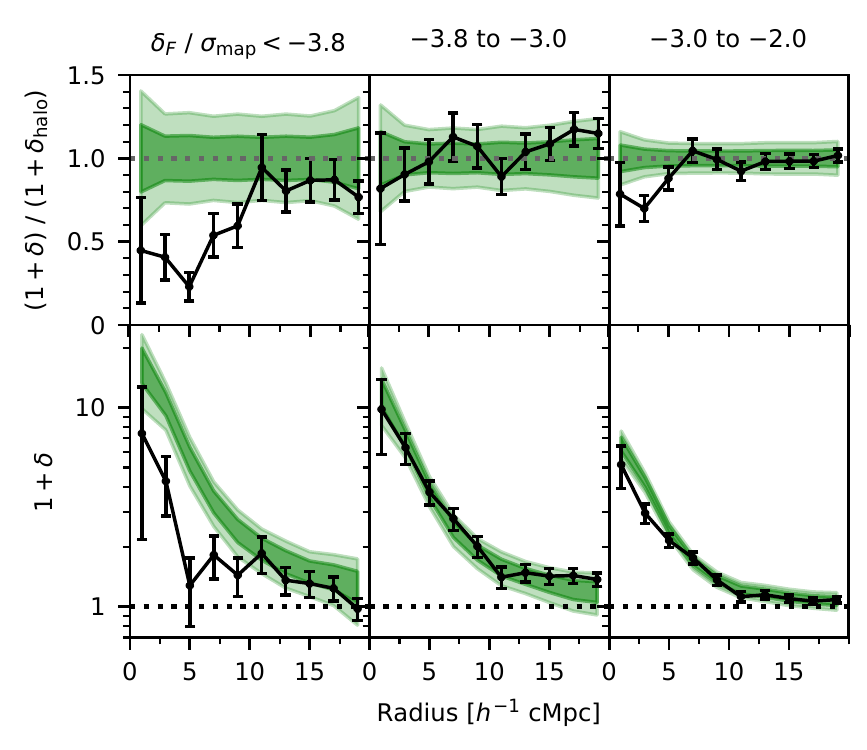}
    \caption{{\bf Average radial profiles of the galaxy or halo overdensity around Ly$\alpha$ absorption peaks.} The observed $1+\delta_{\rm gal}$ (black points with 1 s.e.~Poisson uncertainties) and simulated $1+\delta_{\rm halo}$ (green bands enclosing 68\% and 95\% of mock surveys with Poisson noise subtracted; Methods) are measured differentially in a series of nested cylinders, centered on absorption peaks grouped into three bins of $\delta_F$. In the top row, both curves are divided by the expected $1 + \delta_{\rm halo}$. The quantity $(1 + \delta_{\rm gal}) / (1 + \delta_{\rm halo})$ is proportional to the fraction of halos that host UV-selected galaxies, which declines markedly toward the center of the strongest absorption peaks.}
    \label{fig:radial}
\end{figure}

The only prior survey to systematically examine the galaxy contents of Ly$\alpha$ absorption-selected protoclusters reported high galaxy overdensities in three systems.\citep{Cai17,Shi21} While this may seem to contradict our findings, we emphasize that in most protoclusters, we do measure a galaxy overdensity in accord with simple expectations. Evidence for an UV-dim population only emerges with a larger sample. Further, both the Ly$\alpha$ absorption and the galaxy overdensity were measured quite differently in the earlier studies: the absorption was measured in 4 sightlines across 20 $h^{-1}$ cMpc, a areal density $15\times$ lower than LATIS, and different tracers (Ly$\alpha$ and H$\alpha$ emitters) were used to measure $\delta_{\rm gal}$. The latter may prove to be an illuminating distinction.

Despite many searches, a significant population of massive protoclusters at $z \sim 2.5$ seems to have been missed because their galaxy populations are UV-dim. Studies of galaxy evolution in various $z \sim 2$-3 environments may have omitted the structures in which the environmental influence is strongest. A multiwavelength spectroscopic campaign is needed to conduct a galaxy census in the LATIS absorption peaks, measure the galaxy overdensity via multiple tracers, and understand the unusual galaxy properties. If a comprehensive search does not find substantial galaxy overdensities, it would suggest a surprising break in the connection between Ly$\alpha$ absorption and the density field on scales of several cMpc, a cornerstone of modern cosmology that is thought to be well understood.

\methods

~\\
{\bf Tomographic maps.} We created tomographic maps using the spectra obtained as part of LATIS between Dec.~2017 and Apr.~2021. The survey and map making procedures have been fully described.\citep{Newman20}  The updated data set used here comprises observations of 11 LATIS ``footprints,'' which are distributed among the COSMOS and Canada--France--Hawaii Legacy Survey (CFHTLS) D1 and D4 fields. Only one planned footprint (D4M1, ref.~\citep{Newman20}) has not been observed. In two footprints (D1M1 and D1M2), only half of the planned targets have been observed. We took this into account throughout our analysis, but we note that none of the strongest Ly$\alpha$ absorption peaks of main interest are located in these footprints. We used a total of 2,596 sightlines to construct the Ly$\alpha$ tomographic maps. The maps span a total area of 1.5 deg${}^2$ and a volume, extending from $z=2.2$ to 2.8, of $3.7 \times 10^6 h^{-3}$~cMpc${}^3$.
The $\delta_F$ maps are smoothed by an isotropic gaussian kernel having $\sigma_{\rm kern} = 4$ $h^{-1}$~cMpc. They are then normalized by their standard deviation $\sigma_{\rm map}$, which is consistent among fields (ranging from 0.0457 to 0.0466) and with the mock surveys. Absorption peaks are identified as local minima. We exclude minima within $\sigma_{\rm kern}$ of the map boundary to suppress edge effects. We also iteratively remove minima located within 8~$h^{-1}$~cMpc of another minimum, beginning with the shallowest. Though not essential, this step ensures that the measurements are nearly statistically independent. 

\noindent {\bf Galaxy density measurements.} We measured galaxy overdensities using 2,241 redshifts of Lyman-break galaxies (LBGs) and quasars located within the tomographic map volume. Only high-confidence (${\tt zqual} = 3$ or 4, ref.~\citep{Newman20}) LATIS redshifts were used. The accuracy of the systemic redshifts of the LBGs was verified based on the symmetry of the average Ly$\alpha$ absorption in transverse sightlines.\citep{Rakic11} 
At the location of each absorption peak, we counted the number $N$ of galaxies within a cylinder that has a transverse radius $R$ and extends $\pm R$ along the line of sight. Redshift was converted to line-of-sight distance assuming a pure Hubble flow, so like the $\delta_F$ maps, galaxy densities are measured in velocity space. In Fig.~2 we count galaxies within a $R = 8$ $h^{-1}$~cMpc cylinder, which optimizes the signal-to-noise ratio in $\delta_{\rm gal}$ based on the simulations outlined below. In Fig.~3 we count galaxies differentially in a series of nested cylinders, which are labeled by the mean radius $R$ of the inner and outer cylinders. The half-length of the cylinder along the line of sight is always equal to the radius. The mean galaxy space density $\overline{N}$ was evaluated within each survey field and was found not to vary systematically with redshift within the map volume. The galaxy overdensity is then $\delta_{\rm gal} = N / (\overline{N} V) - 1$, where $V$ is the search volume excluding unobserved portions (i.e., those outside the survey footprint or close to bright stars). The space density of LATIS galaxies is $\overline{N} = 7 \times 10^{-4}$ $h^3$ cMpc${}^{-3}$ corresponding to an average $\overline{N} V = 2.4$ galaxies in randomly placed cylinders.

This simple procedure is sufficient because of the uniformity of LATIS over the relevant angular scales. The fraction of candidates that were observed and assigned a confident redshift varies within each field with an rms of 10\%, when measured in apertures of size $R = 8$ $h^{-1}$~cMpc~$\sim 6.8'$ and accounting for the survey mask. These variations are $5\times$ smaller than the typical uncertainty in $\delta_{\rm gal}$ and are not correlated with $\delta_{\rm gal}$, so they are safely neglected. Geometric constraints on slitmask design, which arise from the requirement that galaxies' spectra not overlap, induce only a negligible bias in $\delta_{\rm gal}$. The fraction of target pairs that were observed is constant for angular separations larger than 12 arcsec, much smaller than any aperture considered in this paper. To verify the insensitivity of $\delta_{\rm gal}$ to slit density limits, we selected halos from a large $N$-body simulation (see below), added a randomly distributed population of foreground interlopers, and placed the combined ``targets'' onto two large pseudo-masks (equivalent to 18 deg${}^2$) respecting mask design constraints. We find that the recovered halo space densities, for the range of apertures and overdensities shown in Fig.~3, are biased by $\lesssim 5\%$. Finally, although the target selection probability is dependent on $r$-band magnitude,\citep{Newman20} its distribution is consistent for galaxies located within absorption peaks binned according to $\delta_F$; thus, trends in target selection probability cannot affect trends with $\delta_F$ (Figs.~2 and 3).

\noindent {\bf Galaxy clustering.} We measured the autocorrelation function $w_p(R)$ of LATIS galaxies at $2 < z < 3$, projected within a velocity window corresponding to $\pm 20$ $h^{-1}$ cMpc, using the Landy-Szalay estimator.\citep{LS} The results are shown for the COSMOS and CFHTLS-D1 fields in Extended Data Fig.~\ref{fig:clustering}. (The D4 field is too small to usefully estimate $w_p$.) Random pairs were generated respecting the survey mask and redshift distribution. The integral constraint, which accounts for the survey geometry, was evaluated following standard methods. \citep{Roche02}  We then computed $w_p(R)$ for dark matter subhalos in the MultiDark {\it Planck} 2 (MDPL2) $N$-body simulation\citep{Klypin16} using the snapshot at $z=2.535$. We considered halos with virial masses $M_{\rm vir} >M_{\rm min}$ for several values of $M_{\rm min}$. We fit the LATIS measurements to the halo clustering over the range $1 < R / (h^{-1}~{\rm cMpc}) < 7$ and derive $M_{\rm min}/{\rm M}_{\odot} = 10^{11.89\pm0.03}$ and $10^{11.70\pm0.05}$ for the COSMOS and D1 fields, respectively. Since the field-to-field difference is larger than the formal uncertainties, 
 we average the two fields and consider halos with $M_{\rm vir} > 10^{11.8} M_{\odot}$ by default, consistent with ref.~\citep{Trainor12}.
 
We tested the sensitivity of our results to the galaxy-halo connection. First, we repeated the analysis perturbing the halo mass limit by the field-to-field variation of $\pm 0.1$~dex and found that the small $p$-values derived from Fig.~2 changed by less than a factor of 2. Second, we considered a model in which the probability of a halo hosting a LATIS-selected galaxy varies smoothly with mass. This probability followed the error function commonly used in halo occupation distribution models, which is 0.5 at a mass $M_{\rm min}$ and transitions from 0 to 1 at a rate controlled by $\sigma_{\rm log M}$ (see eqn.~6 of ref.~\citep{Durkalec18}). We set $\sigma_{\log M} = 0.6$ dex to match the highest values inferred from LBG clustering\citep{Ishikawa17,Durkalec18} and then tuned $M_{\rm min}$ to match our clustering data. In this model, LATIS galaxies occupy halos with a wide mass range ($\log M_{\rm halo}/{\rm M}_{\odot}$ has a median of 12.0 and a s.d.~of 0.45) compared to our fiducial mass-limited model. We find that the comparison between the observed and mock trendlines in Fig.~2 is unaffected. The reason is that the center of the $\delta_{\rm halo}$ trendline distribution is fixed by the clustering data while its width is dominated by observational errors, not scatter in the halo bias.

\noindent {\bf Mock surveys.} We created a suite of mock LATIS surveys within the 1 Gpc${}^3$ MDPL2 simulation. The flux field at $z=2.535$, near the LATIS midpoint, was calculated using the fluctuating Gunn--Peterson approximation,\citep{Gunn65,Weinberg97} which we previously found to accurately reproduce the distribution of fluctuations in the observed maps.\citep{Newman20} For each mock survey realization and each of the three LATIS survey fields, we sampled the simulated flux field using the exact configuration of the observed sightlines (i.e., the relative positions of each spectral pixel in $x$, $y$, and $z$). The mock spectra were smoothed and resampled to match the observations. Random noise and continuum errors were injected following the noise properties of each sightline. We applied mean flux regularization\citep{Lee12} and masked lines with high equivalent width\citep{Newman20} to suppress damped absorbers, following the same methods applied to the observations. A mock tomographic map was created using the {\tt dachshund} Wiener filtering code\citep{Stark15} using the same parameters applied to the observations. The process was repeated over 100 mock surveys covering nearly disjoint subvolumes of the simulation. To calculate the halo overdensity at a simulated absorption peak, we selected a random subset of subhalos in the search aperture having $M_{\rm vir} > 10^{11.8} M_{\odot}$ (see ``Galaxy clustering''), with the selection probability set to match the global space density of halos and LATIS galaxies. We propagated the survey mask into the simulated volumes and excluded halos in ``unobserved'' regions. Extended Data Figure~\ref{fig:marginals} shows that the mock surveys match the observed distributions of $\delta_F$ and $\delta_{\rm gal}$ with impressive accuracy.

We additionally created a separate suite of mock surveys using the IllustrisTNG300-1 simulation.\citep{Nelson19} Although its (205 $h^{-1}$ cMpc)${}^3$ volume is only about twice that of LATIS, and thus too small to capture variance in the density field itself, these mocks are useful to test the relevance of physics not present in the collsionless MDPL2 simulation, including hydrodynamics, feedback, and the formation of high-column density (HCD) lines from dense gas. We used the {\tt fake\_spectra} code\citep{Bird15,Bird17} to create 168,000 high-resolution synthetic spectra on a regular grid in the $z=2.58$ snapshot. Self-shielding of dense gas was modeled using results from radiative transfer calculations \citep{Rahmati13}; other details may be found in ref.~\citep{Qezlou21}. From the TNG spectra, we created 100 mock surveys. In each realization, we selected random sightlines with a density equal to LATIS, averaged over the redshift range of the maps (750 deg${}^{-2}$), and we assigned each sightline a continuum-to-noise ratio drawn from the observed distribution. Mock observations were then synthesized from the spectra as described above. One physical effect not present in TNG is enhanced photo-ionization near quasars. This is expected to have a mild effect on the protocluster signal in most cases.\citep{Miller21} More importantly, it can only reduce the observed absorption, and thus cannot account for our main result, the low galaxy density observed around the strongest absorption peaks. The same comment applies to any  ``pre-heating'' of the intracluster medium\citep{Kooistra21} beyond the feedback implemented in TNG.

\noindent {\bf The $\delta_F-\delta_{\rm gal/halo}$ connection.} To produce the distribution of trendlines in Fig.~2, we sampled random subsets of absorption peaks from the full suite of MDPL2 mock surveys. The subsets were constrained to have the same number of peaks in several bins of $\delta_F$ as the LATIS data, in order to correctly model the uncertainty in the mean $\langle \delta_{\rm gal} \rangle$ conditional on $\delta_F$. For each such subset, we calculated a non-parametric trendline using locally weighted scatterplot smoothing (LOESS\citep{Cappellari13}), a form of local linear regression. The same method was applied to the observations. Shaded regions in Fig.~2 show percentiles of the suite of trendlines. The percentile curves are non-monotonic because the uncertainty increases rapidly at low $\delta_F$ due to the declining number of absorption peaks. Varying the method used to define the trendlines (e.g., varying the fraction of datapoints used in the local fitting between 30\%-70\%, or using a second-order global polynomial fit instead of LOESS) affects the trendline shape but has a minor effect (0.1 s.d.) on the location of the observed trendline within the mock distribution. Excluding the lowest-$\delta_F$ data point is similarly inconsequential. The mock surveys based on MDPL2 (used in Figs.~2 and 3) and TNG produce very similar trendline distributions (Extended Data Fig.~\ref{fig:compare_mocks}a), despite substantial differences in the physics and the mock observation synthesis. We thus find that a simple treatment of the IGM physics is sufficient for our purposes. 

The distributions of $\delta_{\rm gal}$ and $\delta_{\rm halo}$ for individual absorption peaks, in several bins of $\delta_F$, are shown in Extended Data Fig.~\ref{fig:dgalhist}. Generalizing the result in Fig.~2, we find that the observed and mock survey distributions, not only their means, are consistent in all but the strongest absorption peaks. To construct the $\delta_{\rm halo}(R)$ profiles in Fig.~3 from the mock surveys, we subtracted the Poisson noise in quadrature from the standard deviation of $\delta_{\rm halo}$ so that the width of the bands indicates the additional sources of error (e.g., bin-to-bin scatter from errors in $\delta_F$) that are hard to estimate otherwise. The standard Poisson noise is instead shown as error bars on the observations. The mock surveys naturally incorporate the 4 $h^{-1}$ cMpc (1 s.e.) uncertainty in the absorption peak position, which depresses $\delta_{\rm gal}$ and $\delta_{\rm halo}$ in the inner bins.

\noindent {\bf High-column density absorption lines.} HCD lines, including Lyman-limit systems and damped Ly$\alpha$ systems (DLAs) that occur near galaxies, produce absorption that could wrongly be interpreted as a large-scale matter overdensity if it were assumed to arise from the diffuse IGM. Contamination by HCD lines is the predominant concern for single-sightline surveys\citep{Cai16} but is expected to be much reduced in tomographic surveys like LATIS, in which each map resolution element is probed by $\sim10$ sightlines.

We filter lines with high equivalent width to mask DLAs, and the same procedure\citep{Newman20} is applied to the observations and all mock surveys. We particularly examined the 77 sightlines near the strongest absorption peaks (Extended Data Fig.~\ref{fig:sightline_dF}) and found that 88\% have $\delta_F < 0$, indicating that the peaks are produced by widespread IGM absorption, consistent with expectations from the mock surveys with no evidence of contamination by DLAs. More subtle effects were evaluated probabilistically using TNG. Since the trendline distributions in the TNG- and MDPL2-based mocks are very similar (Extended Data Figure~\ref{fig:compare_mocks}a), particularly in the low-$\delta_F$, low-$\delta_{\rm halo}$ region of main interest in this paper, HCD lines seem to negligibly affect the analysis in Fig.~2. To further probe this insensitivity, we precisely isolated the effect of HCD lines by creating a matched suite of TNG mock surveys, which are identical to the fiducial TNG mocks except that all HCD lines ($N_{\rm HI} > 10^{17.2}$~cm${}^{-2}$) are explicitly masked. The differential effect of HCD lines on individual absorption peaks is shown in Extended Data Fig.~\ref{fig:compare_mocks}b. As expected, the distribution has a tail to negative shifts, i.e., enhanced absorption produced by HCD lines. We consider the influence of these shifts on the trendline distribution in Extended Data Fig.~\ref{fig:compare_mocks}c, which shows that excluding HCD lines produces a shift of just 0.06 s.d. Although this may be a slight underestimate, since the frequency of HCD lines in the TNG spectra is $30\pm9\%$ lower than observed\citep{Zafar13} (depending on the column densities considered), the effect of HCD lines would be very minor even if doubled. We conclude that the significance of the discrepancy in Fig.~2 is virtually insensitive to the presence of HCD lines, because the map perturbations they produce are subdominant to other sources of noise in $\delta_F$ and $\delta_{\rm gal}$.

\noindent {\bf Connection to the matter overdensity at $z \sim 2.5$ and massive halos at $z=0$.} The mock surveys allow us to connect the observed $\delta_F$ to the density contrast $\delta_m = \rho / \langle \rho \rangle - 1$ at the observed epoch, smoothed on $\sigma = 4$ $h^{-1}$ cMpc scales like the flux field. The left panel of Extended Data Fig.~\ref{fig:massdist} shows that the absorption peaks correspond to progressively larger matter overdensities as $\delta_F$ decreases, reaching $4.6\sigma$ overdensities, on average, when $\delta_{\rm F}/\sigma_{\rm map} < -4$. 
 We also evaluated the connection to massive halos at $z=0$ by finding the most massive halo within 4 $ h^{-1}$ cMpc of each absorption peak and tracing the mass of its descendant, similar to earlier work.\citep{Stark15} The right panel of Fig.~\ref{fig:massdist} shows that the distributions overlap substantially, but the strongest absorption peaks are almost always associated with protoclusters ($M_{\rm vir}(z=0) / {\rm M}_{\odot} > 10^{14}$). 

\noindent {\bf Environmental trends in the halo mass distribution.} The halo mass distribution is expected to be tilted to higher masses in dense environments traced by strong Ly$\alpha$ absorption.\citep{Lee16B} LATIS galaxies span about 1.3 dex in stellar mass\citep{Weaver22}, which is thought to correspond to only $\sim0.6$ dex in halo mass\citep{Behroozi13}. We find that the shape of the halo mass distribution, measured in our mock surveys within subvolumes defined by $\delta_F$, varies only slightly over this narrow mass range: even considering all halos with $M_{\rm vir} > 10^{11.8} M_{\odot}$, the median $M_{\rm vir}$ shifts by $<0.1$ dex within the strongest absorption peaks. We conclude that higher halo masses alone are unlikely to account for a major shift in the properties of galaxies within absorption peaks.

\noindent {\bf Photometric redshift maps.} We evaluated the overdensity of near-infrared-selected galaxies ($K_s < 25$ AB) around the LATIS absorption peaks located in the COSMOS field using a photometric redshift catalog.\citep{Weaver22} By comparison to the LATIS spectroscopic redshifts, we found that the $z_{\rm phot}$ estimates have a random uncertainty of $\sigma_{\rm zphot} / (1+z) = 0.03$ and a redshift-dependent bias reaching up to $\Delta z / (1+z) = 0.03$. We corrected for this bias in our comparisons. We calculated $\delta_{\rm gal}$ within a redshift interval of $\pm \sigma_{\rm zphot} / 2$ centered on each LATIS absorption peak, following an earlier $z_{\rm phot}$-based protocluster search\citep{Chiang14}, and within a transverse separation of 8 $h^{-1}$ cMpc to match Fig.~2. We simulated the same observation in our mock surveys by considering halos with $M_{\rm vir} > 10^{11.4} {\rm M}_{\odot}$, which match the space density of the $z_{\rm phot}$ sample, and perturbing their $z$ coordinates according to the $z_{\rm phot}$ uncertainty. Extended Data Figure~\ref{fig:zphot} shows that the overdensities are greatly reduced from the spectroscopic $\delta_{\rm gal}$ (Fig.~2) due to the large smearing along the line of sight. Similar to Fig.~2, the observed $\delta_F-\delta_{\rm gal}$ trendline agrees with the mean mock survey at $\delta_F / \sigma_{\rm map} > -3$ and falls to lower $\delta_{\rm gal}$ at more extreme $\delta_F$; unlike Fig.~2, this discrepancy is not significant ($\sim 1.5 \sigma$). We conclude that spectroscopic trend in Fig.~2 is not inconsistent with the photometric redshift analysis, but the large $z_{\rm phot}$ uncertainties prevent a strong test.

\noindent {\bf Counterparts of the strongest absorption peaks in the literature.} In Extended Data Table~\ref{tab:coords} we provide the coordinates and properties of the 8 strongest LATIS absorption peaks ($\delta_F/\sigma_{\rm map} < -3.8$) discussed in this paper. Among these, LATIS1-D2-3 is part of a large proto-supercluster known as Hyperion, \citep{Cucciati18} of which components have previously been identified using galaxy surveys\citep{Diener15,Chiang15,Casey15} and Ly$\alpha$ tomography.\citep{Lee16} LATIS1-D2-1 was previously identified (ID 20) in a catalog of galaxy overdensities identified using photometric redshifts;\citep{Chiang14} it is the highest such overdensity in the portion of the catalog that overlaps the LATIS maps, and it has the highest $\delta_{\rm gal}$ listed in Extended Data Table~1. LATIS1-D2-4 has a potential match (ID 25) in the same catalog, but we do not see a corresponding structure in our photometric redshift maps, which we calibrated against a large spectroscopic library, and conclude that this match is uncertain. As far as we are aware, the other structures have not been previously discussed.

\noindent{\bf Tests of sky subtraction.} Since the flux density of the Ly$\alpha$ forest is typically $30 \times$ smaller than the sky background, small biases in sky subtraction could affect the galaxy spectra. Sky subtraction was performed on a per-exposure basis using two-pass, iterative b-spline modeling of the two-dimensional spectra. We evaluated biases by stacking the two-dimensional spectra and comparing the target (``on-source'') to a ``blank'' part of the slit (``off-source''), which was taken as the region 1.6-2.4 arcsec distant from the target (the slit length is 6 arcsec). Since we are particularly interested in the sky subtraction quality near the absorption peaks, we considered the sightlines with a transverse distance within 8 $h^{-1}$ cMpc of an absorption peak. (We found consistent results when considering only the strongest absorption peaks.) These spectra were shifted into the rest frame of the absorption peak, continuum normalized, and averaged. The Ly$\alpha$ absorption is clearly evident on source, as expected. We also find some residual light off source; after subtracting it from the on-source spectrum and renormalizing the continuum, the Ly$\alpha$ absorption increases modestly in strength by 5\%. This suggests that there may be a residual additive ``pedestal'' around 5\% of the mean Ly$\alpha$ forest flux, or equivalently 0.2\% of the sky background. If the pedestal comprises residual sky background, we note that its spectrum is essentially smooth at the resolution of the maps. An additive uncertainty of 5\% translates to an equal multiplicative uncertainty in $\delta_F$. This should minimally affect our analysis, since $\delta_F$ is ultimately normalized by $\sigma_{\rm map}$ and so is insensitive to a constant multiplicative rescaling. To verify this we propagated the effect of an additive pedestal through our analysis, assuming that it can be represented as a constant count rate of $6 \times 10^{-5}$ electron s${}^{-1}$ pixel${}^{-1}$, which reproduces the fractional estimate above. We subtracted this rate from all two-dimensional spectra, re-extracted the one-dimensional spectra, and re-created the maps. Comparing the strength of absorption peaks between the modified the fiducial maps, we find that the difference $\Delta \delta_F / \sigma_{\rm map}$ has a mean of 0, as anticipated, and a standard deviation of 0.2, which is $3 \times$ smaller than the random noise and so has a minimal effect. In particular, the trendline in Fig.~2 is not notably affected. We conclude that our results are robust to small systematic uncertainties in the sky subtraction. 

\noindent {\bf Data availability.} Data supporting the findings of this study are available from the corresponding author upon request.

\noindent {\bf Code availability.} We have made use of public codes including {\tt astropy}\citep{Astropy1,Astropy2} and {\tt dachshund}.\citep{Stark15} Other supporting analysis code is available from the corresponding author upon request.

{\noindent {\bf Methods References}}
\begin{thebibliographytwo}
\expandafter\ifx\csname url\endcsname\relax
  \def\url#1{\texttt{#1}}\fi
\expandafter\ifx\csname urlprefix\endcsname\relax\def\urlprefix{URL }\fi
\providecommand{\bibinfo}[2]{#2}
\providecommand{\eprint}[2][]{\url{#2}}
\footnotesize
\bibitem{Rakic11}
\bibinfo{author}{{Rakic}, O.}, \bibinfo{author}{{Schaye}, J.},
  \bibinfo{author}{{Steidel}, C.~C.} \& \bibinfo{author}{{Rudie}, G.~C.}
\newblock \bibinfo{title}{{Calibrating galaxy redshifts using absorption by the
  surrounding intergalactic medium}}.
\newblock \emph{\bibinfo{journal}{\mnras}} \textbf{\bibinfo{volume}{414}},
  \bibinfo{pages}{3265--3271} (\bibinfo{year}{2011}).

\bibitem{LS}
\bibinfo{author}{{Landy}, S.~D.} \& \bibinfo{author}{{Szalay}, A.~S.}
\newblock \bibinfo{title}{{Bias and Variance of Angular Correlation
  Functions}}.
\newblock \emph{\bibinfo{journal}{\apj}} \textbf{\bibinfo{volume}{412}},
  \bibinfo{pages}{64} (\bibinfo{year}{1993}).

\bibitem{Roche02}
\bibinfo{author}{{Roche}, N.~D.}, \bibinfo{author}{{Almaini}, O.},
  \bibinfo{author}{{Dunlop}, J.}, \bibinfo{author}{{Ivison}, R.~J.} \&
  \bibinfo{author}{{Willott}, C.~J.}
\newblock \bibinfo{title}{{The clustering, number counts and morphology of
  extremely red ($R-K > 5$) galaxies to $K \leq 21$}}.
\newblock \emph{\bibinfo{journal}{\mnras}} \textbf{\bibinfo{volume}{337}},
  \bibinfo{pages}{1282--1298} (\bibinfo{year}{2002}).

\bibitem{Trainor12}
\bibinfo{author}{{Trainor}, R.~F.} \& \bibinfo{author}{{Steidel}, C.~C.}
\newblock \bibinfo{title}{{The Halo Masses and Galaxy Environments of
  Hyperluminous QSOs at $z \sim 2.7$ in the Keck Baryonic Structure Survey}}.
\newblock \emph{\bibinfo{journal}{\apj}} \textbf{\bibinfo{volume}{752}},
  \bibinfo{pages}{39} (\bibinfo{year}{2012}).

\bibitem{Durkalec18}
\bibinfo{author}{{Durkalec}, A.} \emph{et~al.}
\newblock \bibinfo{title}{{The VIMOS Ultra Deep Survey. Luminosity and stellar
  mass dependence of galaxy clustering at $z \sim 3$}}.
\newblock \emph{\bibinfo{journal}{\aap}} \textbf{\bibinfo{volume}{612}},
  \bibinfo{pages}{A42} (\bibinfo{year}{2018}).
%\newblock \eprint{1703.02049}.

\bibitem{Ishikawa17}
\bibinfo{author}{{Ishikawa}, S.} \emph{et~al.}
\newblock \bibinfo{title}{{The Galaxy-Halo Connection in High-redshift
  Universe: Details and Evolution of Stellar-to-halo Mass Ratios of Lyman Break
  Galaxies on CFHTLS Deep Fields}}.
\newblock \emph{\bibinfo{journal}{\apj}} \textbf{\bibinfo{volume}{841}},
  \bibinfo{pages}{8} (\bibinfo{year}{2017}).
%\newblock \eprint{1612.06869}.

\bibitem{Gunn65}
\bibinfo{author}{{Gunn}, J.~E.} \& \bibinfo{author}{{Peterson}, B.~A.}
\newblock \bibinfo{title}{{On the Density of Neutral Hydrogen in Intergalactic
  Space.}}
\newblock \emph{\bibinfo{journal}{\apj}} \textbf{\bibinfo{volume}{142}},
  \bibinfo{pages}{1633--1636} (\bibinfo{year}{1965}).

\bibitem{Weinberg97}
\bibinfo{author}{{Weinberg}, D.~H.}, \bibinfo{author}{{Hernsquit}, L.},
  \bibinfo{author}{{Katz}, N.}, \bibinfo{author}{{Croft}, R.} \&
  \bibinfo{author}{{Miralda-Escud{\'e}}, J.}
\newblock \bibinfo{title}{{Hubble Flow Broadening of the Ly$\alpha$ Forest and
  its Implications}}.
\newblock In \bibinfo{editor}{{Petitjean}, P.} \& \bibinfo{editor}{{Charlot},
  S.} (eds.) \emph{\bibinfo{booktitle}{Structure and Evolution of the
  Intergalactic Medium from QSO Absorption Line System}}, \bibinfo{pages}{133}
  (\bibinfo{year}{1997}).

\bibitem{Lee12}
\bibinfo{author}{{Lee}, K.-G.}, \bibinfo{author}{{Suzuki}, N.} \&
  \bibinfo{author}{{Spergel}, D.~N.}
\newblock \bibinfo{title}{{Mean-flux-regulated Principal Component Analysis
  Continuum Fitting of Sloan Digital Sky Survey Ly{\ensuremath{\alpha}} Forest
  Spectra}}.
\newblock \emph{\bibinfo{journal}{\aj}} \textbf{\bibinfo{volume}{143}},
  \bibinfo{pages}{51} (\bibinfo{year}{2012}).

\bibitem{Bird15}
\bibinfo{author}{{Bird}, S.} \emph{et~al.}
\newblock \bibinfo{title}{{Reproducing the kinematics of damped Lyman
  {\ensuremath{\alpha}} systems}}.
\newblock \emph{\bibinfo{journal}{\mnras}} \textbf{\bibinfo{volume}{447}},
  \bibinfo{pages}{1834--1846} (\bibinfo{year}{2015}).

\bibitem{Bird17}
\bibinfo{author}{{Bird}, S.}
\newblock \bibinfo{title}{{FSFE: Fake Spectra Flux Extractor.}}
\newblock \emph{\bibinfo{journal}{Astrophysics Source Code Library}}, 
   ascl:1710.012 (\bibinfo{year}{2017}).

\bibitem{Rahmati13}
\bibinfo{author}{{Rahmati}, A.}, \bibinfo{author}{{Pawlik}, A.~H.},
  \bibinfo{author}{{Rai{\v{c}}evi{\'c}}, M.} \& \bibinfo{author}{{Schaye}, J.}
\newblock \bibinfo{title}{{On the evolution of the H I column density
  distribution in cosmological simulations}}.
\newblock% \emph{\bibinfo{journal}{\mnras}} \textbf{\bibinfo{volume}{430}},
  \bibinfo{pages}{2427--2445} (\bibinfo{year}{2013}).
%\newblock \eprint{1210.7808}.

\bibitem{Qezlou21}
\bibinfo{author}{{Qezlou}, M.}, \bibinfo{author}{{Newman}, A.~B.},
  \bibinfo{author}{{Rudie}, G.~C.} \& \bibinfo{author}{{Bird}, S.}
\newblock \bibinfo{title}{{Characterizing protoclusters and protogroups at
  $\mathrm{z \sim 2.5}$ using Lyman-$\alpha$ Tomography}}.
\newblock %\emph{\bibinfo{journal}{arXiv e-prints}}
  \bibinfo{pages}{arXiv:2112.03930} (\bibinfo{year}{2021}).
%\newblock \eprint{2112.03930}.

\bibitem{Miller21}
\bibinfo{author}{{Miller}, J. S.~A.}, \bibinfo{author}{{Bolton}, J.~S.} \&
  \bibinfo{author}{{Hatch}, N.~A.}
\newblock \bibinfo{title}{{Searching for the shadows of giants - II. The effect
  of local ionization on the Ly$\alpha$ absorption signatures of protoclusters
  at redshift $z \sim 2.4$}}.
\newblock \emph{\bibinfo{journal}{\mnras}} \textbf{\bibinfo{volume}{506}},
  \bibinfo{pages}{6001--6013} (\bibinfo{year}{2021}).

\bibitem{Kooistra21}
\bibinfo{author}{{Kooistra}, R.}, \bibinfo{author}{{Inoue}, S.},
  \bibinfo{author}{{Lee}, K.-G.}, \bibinfo{author}{{Cen}, R.} \&
  \bibinfo{author}{{Yoshida}, N.}
\newblock \bibinfo{title}{{Detecting preheating in proto-clusters with
  Lyman-$\alpha$ Forest Tomography}}.
\newblock% \emph{\bibinfo{journal}{arXiv e-prints}}
  \bibinfo{pages}{arXiv:2109.09954} (\bibinfo{year}{2021}).

\bibitem{Cai16}
\bibinfo{author}{{Cai}, Z.} \emph{et~al.}
\newblock \bibinfo{title}{{Mapping the Most Massive Overdensity Through
  Hydrogen (MAMMOTH) I: Methodology}}.
\newblock \emph{\bibinfo{journal}{\apj}} \textbf{\bibinfo{volume}{833}},
  \bibinfo{pages}{135} (\bibinfo{year}{2016}).

\bibitem{Zafar13}
\bibinfo{author}{{Zafar}, T.} \emph{et~al.}
\newblock \bibinfo{title}{{The ESO UVES advanced data products quasar sample.
  II. Cosmological evolution of the neutral gas mass density}}.
\newblock \emph{\bibinfo{journal}{\aap}} \textbf{\bibinfo{volume}{556}},
  \bibinfo{pages}{A141} (\bibinfo{year}{2013}).
%\newblock \eprint{1307.0602}.

\bibitem{Lee16B}
\bibinfo{author}{{Lee}, K.-G.} \& \bibinfo{author}{{White}, M.}
\newblock \bibinfo{title}{{Revealing the $z \sim 2.5$ Cosmic Web with 3D
  Ly{\ensuremath{\alpha}} Forest Tomography: a Deformation Tensor Approach}}.
\newblock \emph{\bibinfo{journal}{\apj}} \textbf{\bibinfo{volume}{831}},
  \bibinfo{pages}{181} (\bibinfo{year}{2016}).

\bibitem{Behroozi13}
\bibinfo{author}{{Behroozi}, P.~S.}, \bibinfo{author}{{Wechsler}, R.~H.} \&
  \bibinfo{author}{{Conroy}, C.}
\newblock \bibinfo{title}{{The Average Star Formation Histories of Galaxies in
  Dark Matter Halos from $z = 0-8$}}.
\newblock \emph{\bibinfo{journal}{\apj}} \textbf{\bibinfo{volume}{770}},
  \bibinfo{pages}{57} (\bibinfo{year}{2013}).

\bibitem{Astropy1}
\bibinfo{author}{{Astropy Collaboration}} \emph{et~al.}
\newblock \bibinfo{title}{{Astropy: A community Python package for astronomy}}.
\newblock \emph{\bibinfo{journal}{\aap}} \textbf{\bibinfo{volume}{558}},
  \bibinfo{pages}{A33} (\bibinfo{year}{2013}).

\bibitem{Astropy2}
\bibinfo{author}{{Astropy Collaboration}} \emph{et~al.}
\newblock \bibinfo{title}{{The Astropy Project: Building an Open-science
  Project and Status of the v2.0 Core Package}}.
\newblock \emph{\bibinfo{journal}{\aj}} \textbf{\bibinfo{volume}{156}},
  \bibinfo{pages}{123} (\bibinfo{year}{2018}).
\end{thebibliographytwo}

\clearpage

\noindent {\bf Acknowledgments} This paper includes data gathered with the 6.5-meter Magellan Telescopes located at Las Campanas Observatory, Chile. We gratefully acknowledge the support of the Observatory staff. A.B.N., S.B., and B.C.L. acknowledge support from the National Science Foundation under Grant Nos.~2108014, 2107821, and 1908422, respectively. E.C. acknowledges support from ANID project Basal AFB-170002. This work is based in part on observations obtained with MegaPrime / MegaCam, a joint project of CFHT and CEA/IRFU, at the Canada--France--Hawaii Telescope (CFHT) which is operated by the National Research Council (NRC) of Canada, the Institut National des Science de l’Univers of the Centre National de la Recherche Scientifique (CNRS) of France, and the University of Hawaii. Also based in part on data products produced at Terapix available at the Canadian Astronomy Data Centre as part of the Canada--France--Hawaii Telescope Legacy Survey, a collaborative project of NRC and CNRS.

~\\
\noindent {\bf Author contributions} A.B.N., G.C.R., and G.A.B. designed LATIS, obtained telescope access, and, along with V.P. and E.C., conducted the observations. D.D.K. created the data reduction software. A.B.N., M.Q., and S.B. created the mock surveys. A.B.N. processed the observations, created the Ly$\alpha$ and galaxy density maps, and drafted the manuscript. All authors contributed to the interpretation and manuscript preparation.

~\\
\noindent {\bf Competing interests} The authors declare no competing interests.

~\\
\noindent {\bf Additional Information}

~\\
\noindent {\bf Correspondence and requests for materials} should be addressed to Andrew Newman (anewman@carnegiescience.edu).

~\\
\noindent {\bf Reprints and permissions information} is available at www.nature.com/reprints.

\clearpage

\makeatletter
\renewcommand{\fnum@figure}{Extended Data Fig. \thefigure}
\renewcommand{\fnum@table}{Extended Data Table \thetable}
\makeatother
\setcounter{figure}{0}    

\begin{figure*}
    \centering
    \includegraphics[width=3.5in]{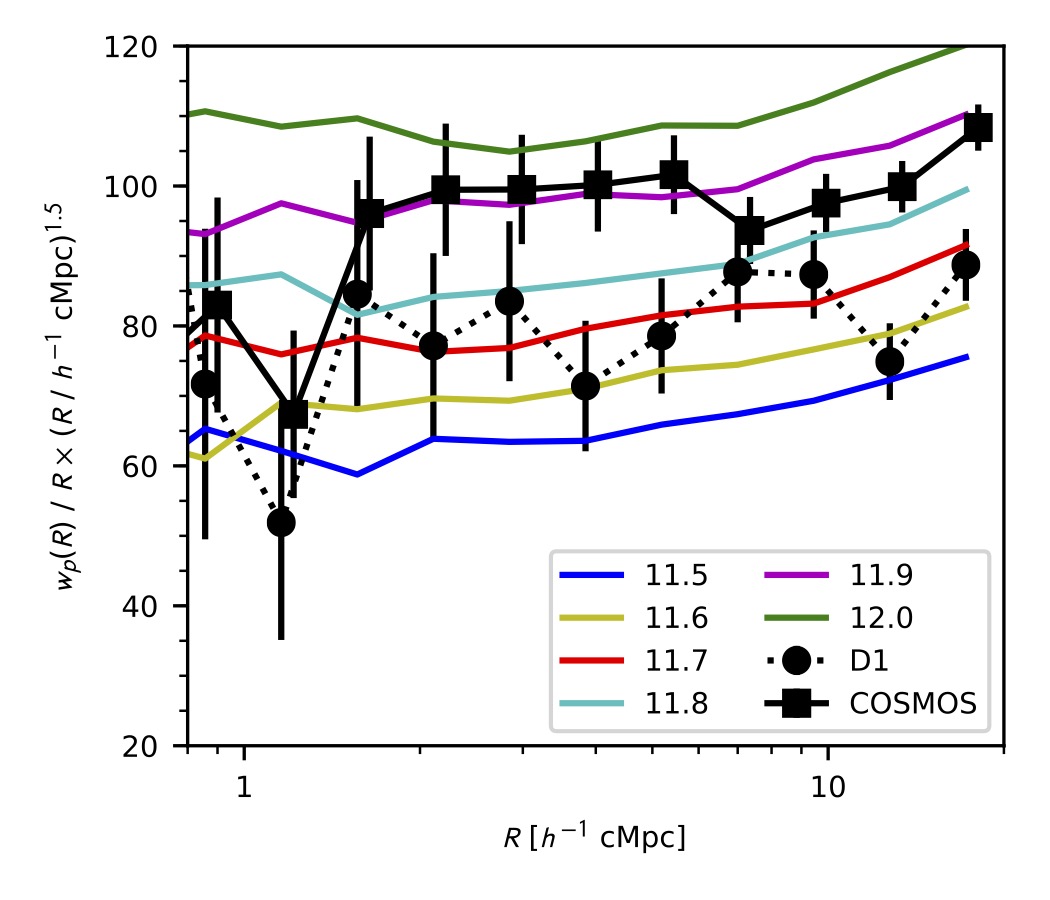}
    \caption{{\bf The projected correlation function $w_p(R)$ of LATIS galaxies.} We consider galaxies at $2 < z < 3$, evaluating $w_p$ separately in the COSMOS and CFHTLS-D1 fields, and compare to the the correlation function of dark matter halos with $M_{\rm vir} > M_{\rm min}$ for several values of $M_{\rm min}$ indicated in the legend. To reduce the range, the dimensionless $w_p(R) / R$ is scaled by $(R / h^{-1} {\rm cMpc})^{1.5}$. Error bars show 1 s.e.~Poisson uncertainties.}
    \label{fig:clustering}
\end{figure*}

\begin{figure*}
    \centering
    \includegraphics[width=7in]{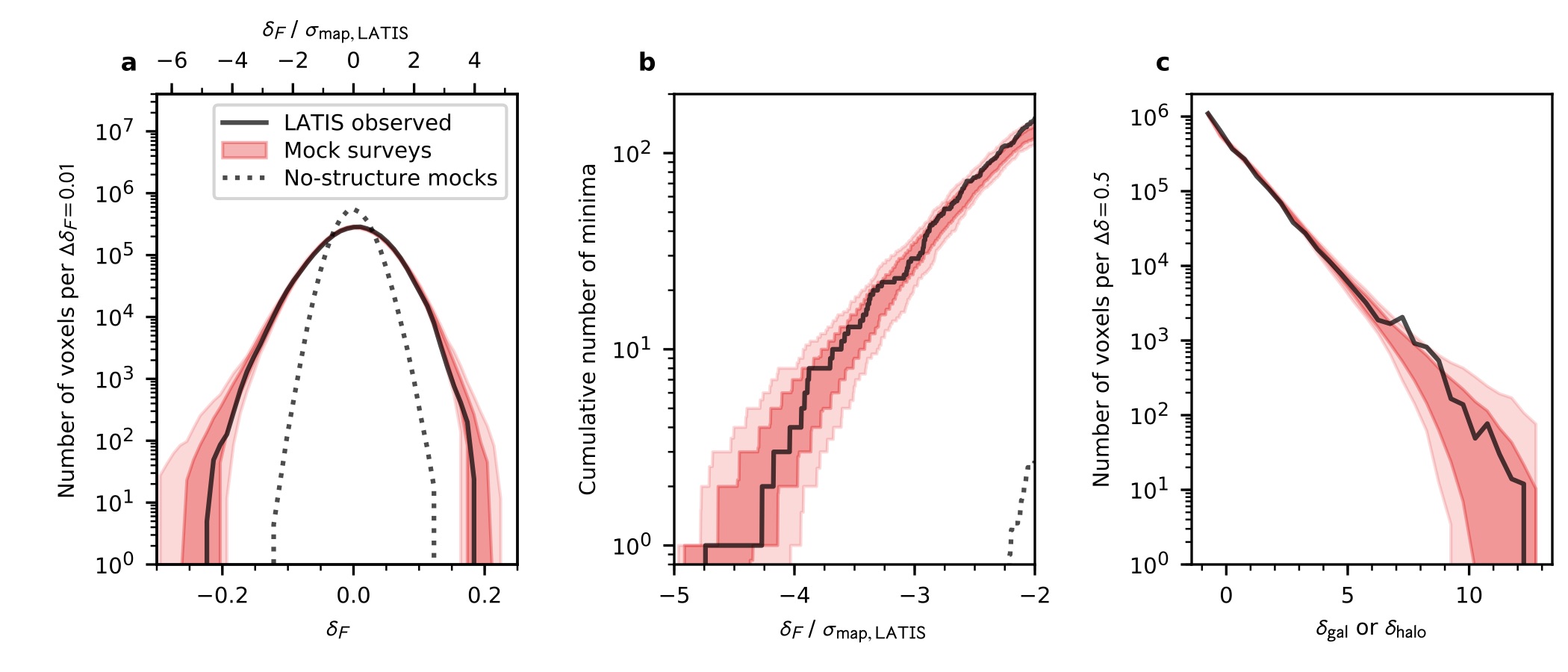}
    \caption{{\bf Comparison of the statistical properties of the LATIS maps and the mock surveys.} {\bf a,} The distributions of the flux contrast $\delta_F$, evaluated at each (1 $h^{-1}$ cMpc)${}^3$ map voxel, in the LATIS maps (black curves) and the MDPL2 mock surveys (red bands enclose 68\% and 95\% of realizations). The dashed curve show the expected distribution from observational noise alone, as estimated from mock surveys of structureless volumes, i.e., with $\delta_F = 0$ everywhere. {\bf b,} The cumulative number of absorption peaks in the LATIS maps and the mock surveys. {\bf c,} The distributions of the galaxy and halo overdensities, evaluated at each map voxel in the observed and mock maps. Altogether the mock surveys accurately match the observed distributions of $\delta_F$ and $\delta_{\rm gal}$ individually.}
    \label{fig:marginals}
\end{figure*}

\begin{figure*}
    \centering
    \includegraphics[width=6.5in]{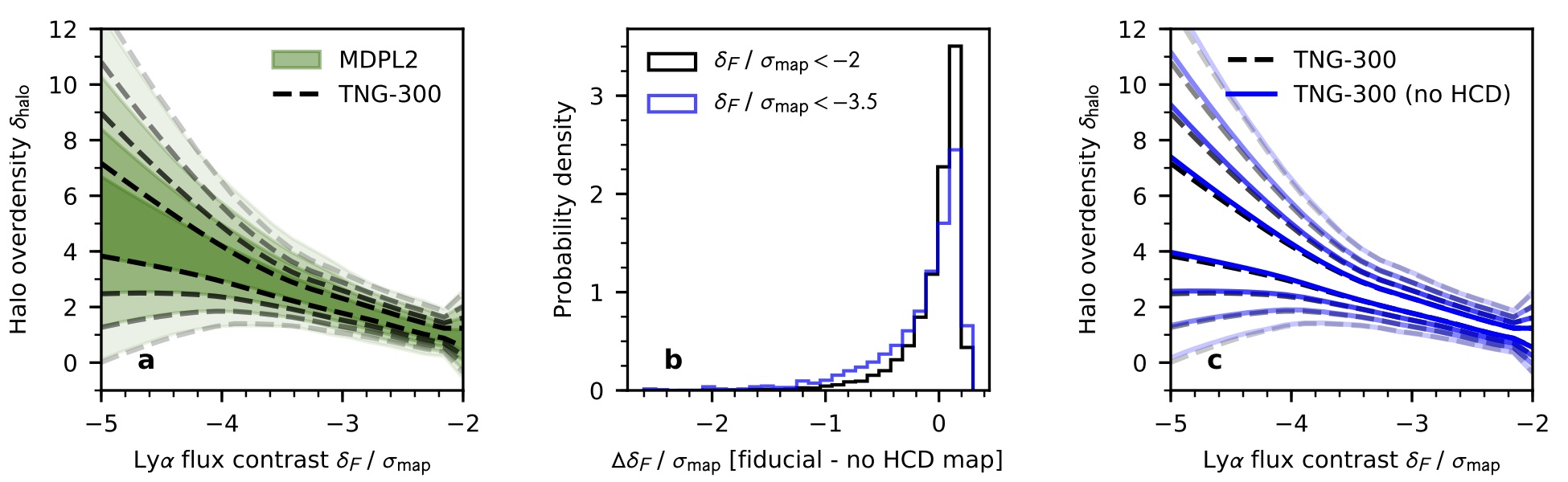}
    \caption{{\bf Evaluation of the robustness of the $\delta_F-\delta_{\rm halo}$ connection.} {\bf a,} Comparison of the trendline distributions, following Fig.~2, derived from mock surveys based on the MDPL2 (dark matter only) and IllustrisTNG300 (including hydrodynamics, galaxy formation, and HCD lines) simulations. For each simulation, curves show percentiles equivalent to 1, 2, 3, and 4 s.d.~in a normal distribution. {\bf b,} The distribution of differences in $\delta_F/\sigma_{\rm map}$ between absorption peaks in the fiducial TNG mock surveys and the corresponding peaks in mocks that explicitly exclude HCD lines ($N > 10^{17.2}$ cm${}^{-2}$). {\bf c,} Comparison of the trendline distributions, following panel {\bf a}, derived from the fiducial TNG mock surveys and those that exclude HCD lines.}
    \label{fig:compare_mocks}
\end{figure*}

\begin{figure*}
    \centering
    \includegraphics[width=7in]{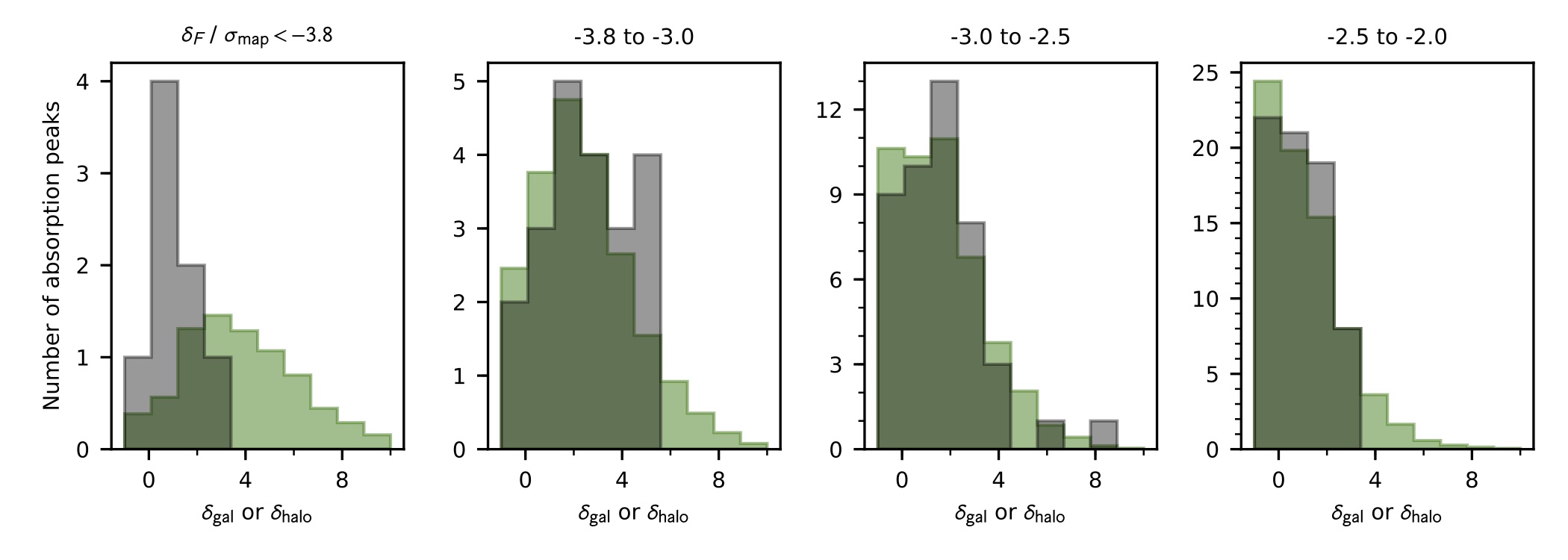}
    \caption{{\bf Distributions of the galaxy or halo overdensity around Ly$\alpha$ absorption peaks.} Gray histograms show 
    $\delta_{\rm gal}$ observed around Ly$\alpha$ absorption peaks within an 8 $h^{-1}$ cMpc aperture following Fig.~\ref{fig:deltas}. Green histograms show the distributions of $\delta_{\rm halo}$ in the MDPL2 mock surveys, normalized to match the integrals of the corresponding gray histograms. The absorption peaks are split into four bins of $\delta_F$, highlighting the discrepancy in the strongest absorption peaks.}
    \label{fig:dgalhist}
\end{figure*}

\begin{figure*}
    \centering
    \includegraphics[width=3.5in]{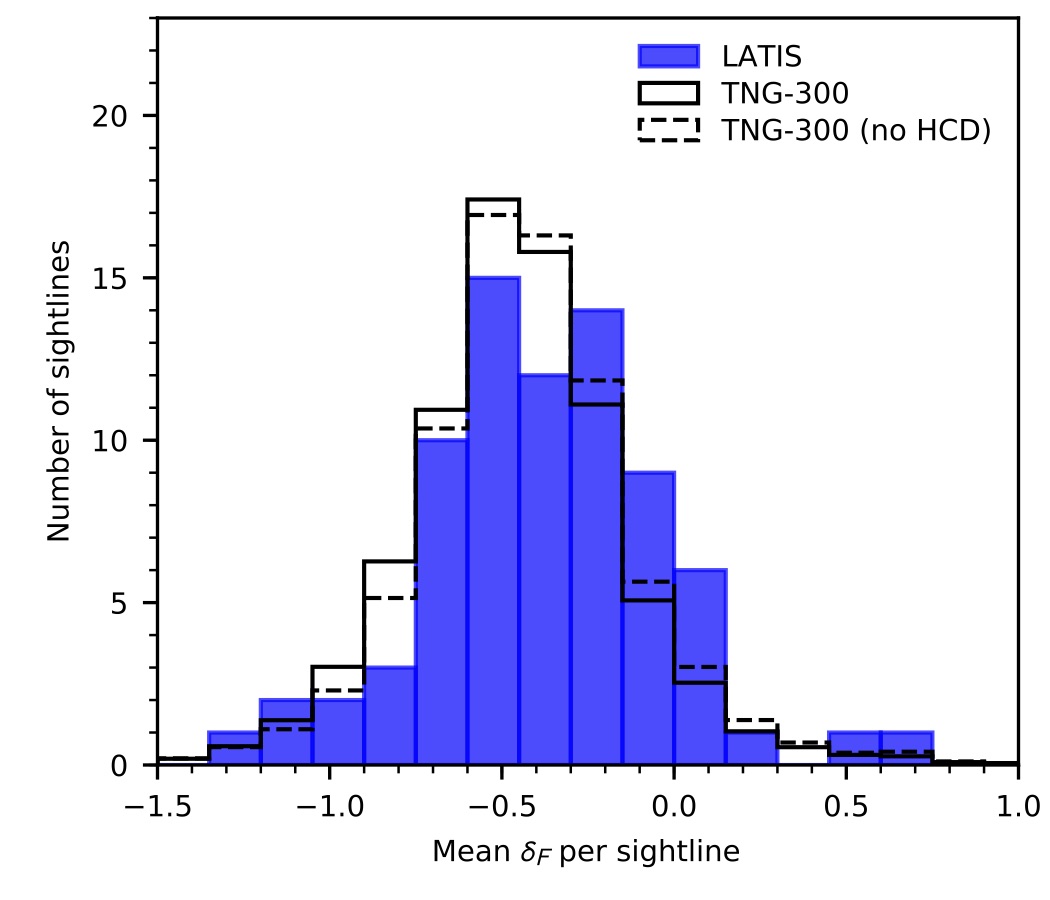}
    \caption{{\bf Evidence of widespread absorption near the strongest absorption peaks.} The distribution of absorption in the 77 sightlines probing the 8 strongest absorption peaks ($\delta_F/\sigma_{\rm map} < -3.8$) within 4 $h^{-1}$ cMpc is shown and compared to the TNG mock surveys. Along each sightline, $\delta_F$ is averaged within $|\Delta z| < 4$ $h^{-1}$ cMpc of the absorption peak. The TNG distributions are scaled to match the integral of the LATIS histogram.}
    \label{fig:sightline_dF}
\end{figure*}

\begin{figure*}
    \centering
    \includegraphics[width=6.5in]{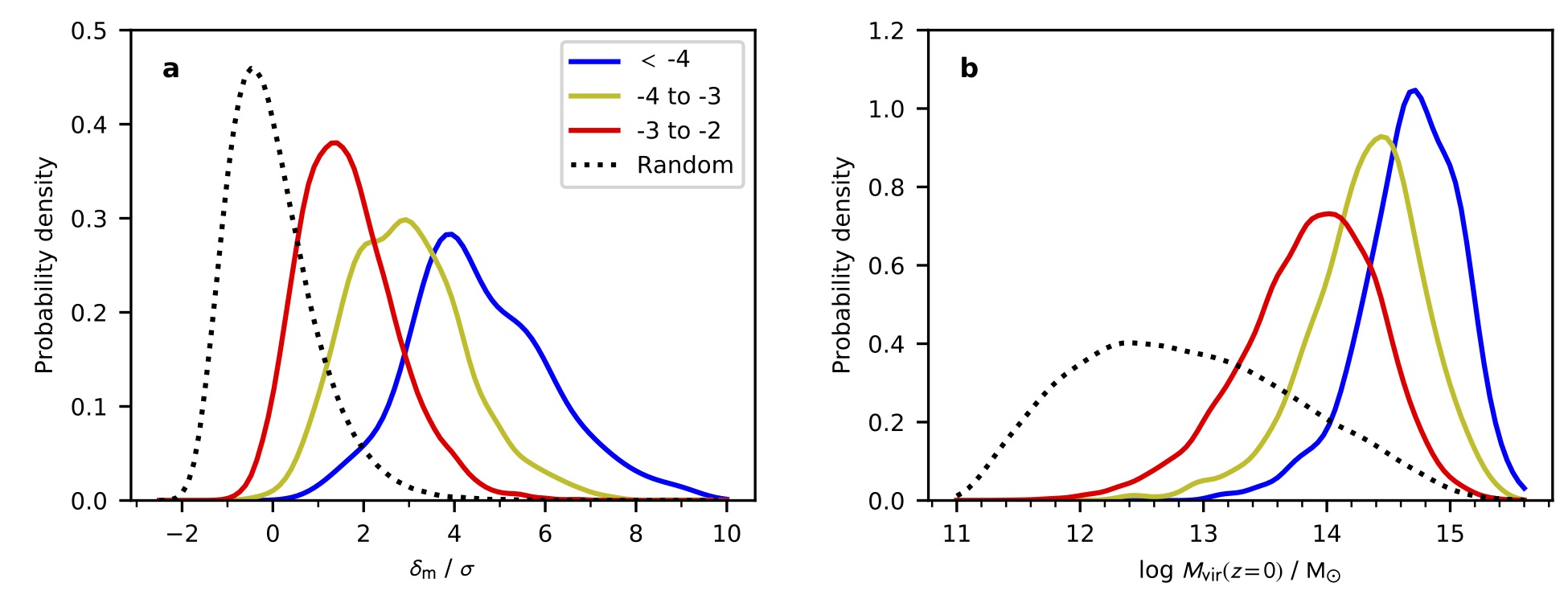}
    \caption{{\bf The connection of absorption peaks with the matter overdensity and descendant halos.} {\bf a,} The distribution of the matter density contrast $\delta_m$, smoothed with a Gaussian kernel having $\sigma = 4$ $h^{-1}$ cMpc and expressed in units of its standard deviation, at the location of absorption peaks in the mock surveys. Bins of $\delta_F/\sigma_{\rm map}$ are indicated in the legend. The dotted curve shows the global distribution. {\bf b,} The analogous distributions of the $z=0$ descendant halo mass $M_{\rm vir}(z=0)$.}
    \label{fig:massdist}
\end{figure*}

\begin{figure*}
    \centering
    \includegraphics[width=3.5in]{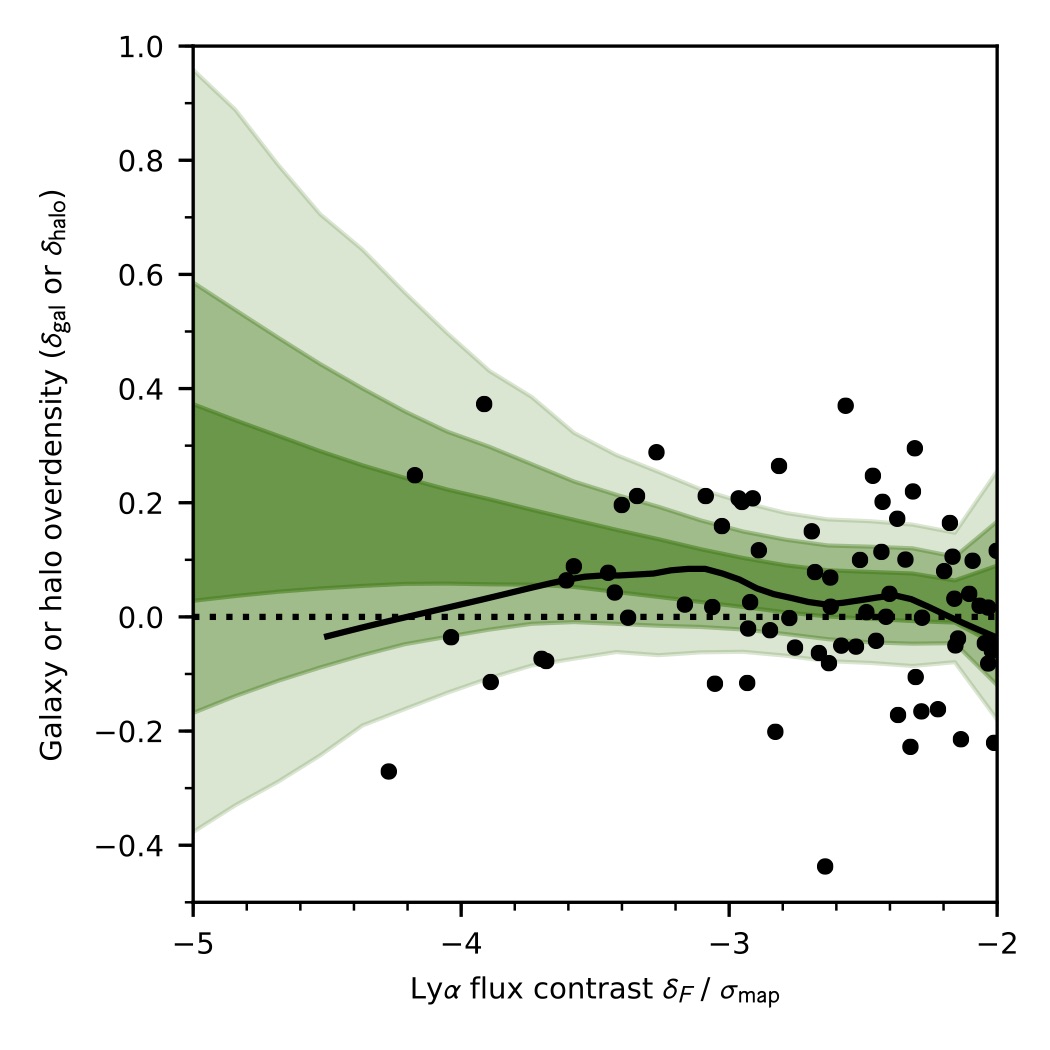}
    \caption{{\bf The $\delta_F-\delta_{\rm gal}$ trend evaluated using photometric redshifts.} For each LATIS absorption peak in the COSMOS field, a black point indicates the galaxy overdensity estimated using photometric redshifts\citep{Weaver22}. A trendline (black curve) analogous to Fig.~2 is overlaid on the distribution of trendlines derived from mock surveys (green), demonstrating the large uncertainties at low $\delta_F$. Bands indicate percentiles equivalent to 1, 2, and 3 s.d.~in a normal distribution.} 
    \label{fig:zphot}
\end{figure*}

\begin{figure*}
\centering

\includegraphics[width=4.4in]{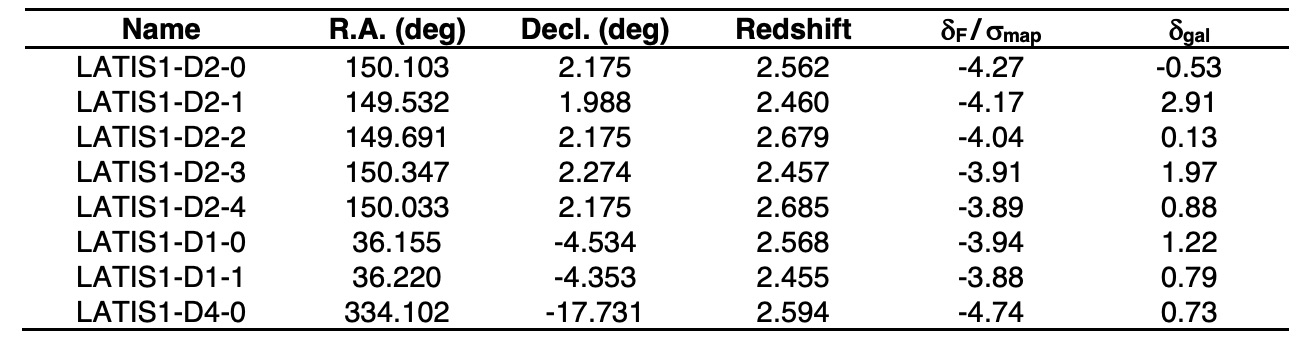}
\caption{{\bf Coordinates and properties of the 8 strongest LATIS absorbers having $\delta_F / \sigma_{\rm map} < -3.8$.} Using the mock surveys, we estimate $1\sigma$ errors of 2.8 arcmin in sky position (per coordinate), 0.002 in redshift, 0.6 in $\delta_F / \sigma_{\rm map}$, and 0.9 in $\delta_{\rm gal}$.}
\label{tab:coords}
\end{figure*}


\clearpage

\begin{thebibliography}{10}
\expandafter\ifx\csname url\endcsname\relax
  \def\url#1{\texttt{#1}}\fi
\expandafter\ifx\csname urlprefix\endcsname\relax\def\urlprefix{URL }\fi
\providecommand{\bibinfo}[2]{#2}
\providecommand{\eprint}[2][]{\url{#2}}
\footnotesize

\bibitem{Overzier16}
\bibinfo{author}{{Overzier}, R.~A.}
\newblock \bibinfo{title}{{The realm of the galaxy protoclusters. A review}}.
\newblock \emph{\bibinfo{journal}{\aapr}} \textbf{\bibinfo{volume}{24}},
  \bibinfo{pages}{14} (\bibinfo{year}{2016}).

\bibitem{Lee14}
\bibinfo{author}{{Lee}, K.-G.}, \bibinfo{author}{{Hennawi}, J.~F.},
  \bibinfo{author}{{White}, M.}, \bibinfo{author}{{Croft}, R. A.~C.} \&
  \bibinfo{author}{{Ozbek}, M.}
\newblock \bibinfo{title}{{Observational Requirements for
  Ly{\ensuremath{\alpha}} Forest Tomographic Mapping of Large-scale Structure
  at $z \sim 2$}}.
\newblock \emph{\bibinfo{journal}{\apj}} \textbf{\bibinfo{volume}{788}},
  \bibinfo{pages}{49} (\bibinfo{year}{2014}).

\bibitem{Newman20}
\bibinfo{author}{{Newman}, A.~B.} \emph{et~al.}
\newblock \bibinfo{title}{{LATIS: The Ly{\ensuremath{\alpha}} Tomography IMACS
  Survey}}.
\newblock \emph{\bibinfo{journal}{\apj}} \textbf{\bibinfo{volume}{891}},
  \bibinfo{pages}{147} (\bibinfo{year}{2020}).

\bibitem{Klypin16}
\bibinfo{author}{{Klypin}, A.}, \bibinfo{author}{{Yepes}, G.},
  \bibinfo{author}{{Gottl{\"o}ber}, S.}, \bibinfo{author}{{Prada}, F.} \&
  \bibinfo{author}{{He{\ss}}, S.}
\newblock \bibinfo{title}{{MultiDark simulations: the story of dark matter halo
  concentrations and density profiles}}.
\newblock \emph{\bibinfo{journal}{\mnras}} \textbf{\bibinfo{volume}{457}},
  \bibinfo{pages}{4340--4359} (\bibinfo{year}{2016}).

\bibitem{Nelson19}
\bibinfo{author}{{Nelson}, D.} \emph{et~al.}
\newblock \bibinfo{title}{{The IllustrisTNG simulations: public data release}}.
\newblock \emph{\bibinfo{journal}{Computational Astrophysics and Cosmology}}
  \textbf{\bibinfo{volume}{6}}, \bibinfo{pages}{2} (\bibinfo{year}{2019}).

\bibitem{Contini16}
\bibinfo{author}{{Contini}, E.}, \bibinfo{author}{{De Lucia}, G.},
  \bibinfo{author}{{Hatch}, N.}, \bibinfo{author}{{Borgani}, S.} \&
  \bibinfo{author}{{Kang}, X.}
\newblock \bibinfo{title}{{Semi-analytic model predictions of the galaxy
  population in protoclusters}}.
\newblock \emph{\bibinfo{journal}{\mnras}} \textbf{\bibinfo{volume}{456}},
  \bibinfo{pages}{1924--1935} (\bibinfo{year}{2016}).

\bibitem{Muldrew18}
\bibinfo{author}{{Muldrew}, S.~I.}, \bibinfo{author}{{Hatch}, N.~A.} \&
  \bibinfo{author}{{Cooke}, E.~A.}
\newblock \bibinfo{title}{{Galaxy evolution in protoclusters}}.
\newblock \emph{\bibinfo{journal}{\mnras}} \textbf{\bibinfo{volume}{473}},
  \bibinfo{pages}{2335--2347} (\bibinfo{year}{2018}).

\bibitem{Pichon01}
\bibinfo{author}{{Pichon}, C.}, \bibinfo{author}{{Vergely}, J.~L.},
  \bibinfo{author}{{Rollinde}, E.}, \bibinfo{author}{{Colombi}, S.} \&
  \bibinfo{author}{{Petitjean}, P.}
\newblock \bibinfo{title}{{Inversion of the Lyman {\ensuremath{\alpha}} forest:
  three-dimensional investigation of the intergalactic medium}}.
\newblock \emph{\bibinfo{journal}{\mnras}} \textbf{\bibinfo{volume}{326}},
  \bibinfo{pages}{597--620} (\bibinfo{year}{2001}).

\bibitem{Stark15}
\bibinfo{author}{{Stark}, C.~W.}, \bibinfo{author}{{White}, M.},
  \bibinfo{author}{{Lee}, K.-G.} \& \bibinfo{author}{{Hennawi}, J.~F.}
\newblock \bibinfo{title}{{Protocluster discovery in tomographic Ly
  {\ensuremath{\alpha}} forest flux maps}}.
\newblock \emph{\bibinfo{journal}{\mnras}} \textbf{\bibinfo{volume}{453}},
  \bibinfo{pages}{311--327} (\bibinfo{year}{2015}).

\bibitem{Lee18}
\bibinfo{author}{{Lee}, K.-G.} \emph{et~al.}
\newblock \bibinfo{title}{{First Data Release of the COSMOS
  Ly{\ensuremath{\alpha}} Mapping and Tomography Observations: 3D
  Ly{\ensuremath{\alpha}} Forest Tomography at $2.05 < z < 2.55$}}.
\newblock \emph{\bibinfo{journal}{\apjs}} \textbf{\bibinfo{volume}{237}},
  \bibinfo{pages}{31} (\bibinfo{year}{2018}).

\bibitem{Lee16}
\bibinfo{author}{{Lee}, K.-G.} \emph{et~al.}
\newblock \bibinfo{title}{{Shadow of a Colossus: A $z = 2.44$ Galaxy
  Protocluster Detected in 3D Ly{\ensuremath{\alpha}} Forest Tomographic
  Mapping of the COSMOS Field}}.
\newblock \emph{\bibinfo{journal}{\apj}} \textbf{\bibinfo{volume}{817}},
  \bibinfo{pages}{160} (\bibinfo{year}{2016}).

\bibitem{Chiang14}
\bibinfo{author}{{Chiang}, Y.-K.}, \bibinfo{author}{{Overzier}, R.} \&
  \bibinfo{author}{{Gebhardt}, K.}
\newblock \bibinfo{title}{{Discovery of a Large Number of Candidate
  Protoclusters Traced by $\sim$15 Mpc-scale Galaxy Overdensities in COSMOS}}.
\newblock \emph{\bibinfo{journal}{\apjl}} \textbf{\bibinfo{volume}{782}},
  \bibinfo{pages}{L3} (\bibinfo{year}{2014}).

\bibitem{Chiang15}
\bibinfo{author}{{Chiang}, Y.-K.} \emph{et~al.}
\newblock \bibinfo{title}{{Surveying Galaxy Proto-clusters in Emission: A
  Large-scale Structure at z = 2.44 and the Outlook for HETDEX}}.
\newblock \emph{\bibinfo{journal}{\apj}} \textbf{\bibinfo{volume}{808}},
  \bibinfo{pages}{37} (\bibinfo{year}{2015}).
%\newblock \eprint{1505.03877}.

\bibitem{Diener15}
\bibinfo{author}{{Diener}, C.} \emph{et~al.}
\newblock \bibinfo{title}{{A Protocluster at $z = 2.45$}}.
\newblock \emph{\bibinfo{journal}{\apj}} \textbf{\bibinfo{volume}{802}},
  \bibinfo{pages}{31} (\bibinfo{year}{2015}).

\bibitem{Casey15}
\bibinfo{author}{{Casey}, C.~M.} \emph{et~al.}
\newblock \bibinfo{title}{{A Massive, Distant Proto-cluster at $z = 2.47$
  Caught in a Phase of Rapid Formation?}}
\newblock \emph{\bibinfo{journal}{\apjl}} \textbf{\bibinfo{volume}{808}},
  \bibinfo{pages}{L33} (\bibinfo{year}{2015}).

\bibitem{Cucciati18}
\bibinfo{author}{{Cucciati}, O.} \emph{et~al.}
\newblock \bibinfo{title}{{The progeny of a cosmic titan: a massive
  multi-component proto-supercluster in formation at $z = 2.45$ in VUDS}}.
\newblock \emph{\bibinfo{journal}{\aap}} \textbf{\bibinfo{volume}{619}},
  \bibinfo{pages}{A49} (\bibinfo{year}{2018}).

\bibitem{Cucciati14}
\bibinfo{author}{{Cucciati}, O.} \emph{et~al.}
\newblock \bibinfo{title}{{Discovery of a rich proto-cluster at $z = 2.9$ and
  associated diffuse cold gas in the VIMOS Ultra-Deep Survey (VUDS)}}.
\newblock \emph{\bibinfo{journal}{\aap}} \textbf{\bibinfo{volume}{570}},
  \bibinfo{pages}{A16} (\bibinfo{year}{2014}).

\bibitem{Lemaux14}
\bibinfo{author}{{Lemaux}, B.~C.} \emph{et~al.}
\newblock \bibinfo{title}{{VIMOS Ultra-Deep Survey (VUDS): Witnessing the
  assembly of a massive cluster at $z \sim 3.3$}}.
\newblock \emph{\bibinfo{journal}{\aap}} \textbf{\bibinfo{volume}{572}},
  \bibinfo{pages}{A41} (\bibinfo{year}{2014}).

\bibitem{Blanton09}
\bibinfo{author}{{Blanton}, M.~R.} \& \bibinfo{author}{{Moustakas}, J.}
\newblock \bibinfo{title}{{Physical Properties and Environments of Nearby
  Galaxies}}.
\newblock \emph{\bibinfo{journal}{\araa}} \textbf{\bibinfo{volume}{47}},
  \bibinfo{pages}{159--210} (\bibinfo{year}{2009}).

\bibitem{Muzzin12}
\bibinfo{author}{{Muzzin}, A.} \emph{et~al.}
\newblock \bibinfo{title}{{The Gemini Cluster Astrophysics Spectroscopic Survey
  (GCLASS): The Role of Environment and Self-regulation in Galaxy Evolution at
  $z \sim 1$}}.
\newblock \emph{\bibinfo{journal}{\apj}} \textbf{\bibinfo{volume}{746}},
  \bibinfo{pages}{188} (\bibinfo{year}{2012}).

\bibitem{Cappellari13}
\bibinfo{author}{{Cappellari}, M.} \emph{et~al.}
\newblock \bibinfo{title}{{The ATLAS$^{\rm 3D}$ project - XX. Mass-size and
  mass-{\ensuremath{\sigma}} distributions of early-type galaxies: bulge
  fraction drives kinematics, mass-to-light ratio, molecular gas fraction and
  stellar initial mass function}}.
\newblock \emph{\bibinfo{journal}{\mnras}} \textbf{\bibinfo{volume}{432}},
  \bibinfo{pages}{1862--1893} (\bibinfo{year}{2013}).

\bibitem{Casey16}
\bibinfo{author}{{Casey}, C.~M.}
\newblock \bibinfo{title}{{The Ubiquity of Coeval Starbursts in Massive Galaxy
  Cluster Progenitors}}.
\newblock \emph{\bibinfo{journal}{\apj}} \textbf{\bibinfo{volume}{824}},
  \bibinfo{pages}{36} (\bibinfo{year}{2016}).

\bibitem{Wang19}
\bibinfo{author}{{Wang}, T.} \emph{et~al.}
\newblock \bibinfo{title}{{A dominant population of optically invisible massive
  galaxies in the early Universe}}.
\newblock \emph{\bibinfo{journal}{\nat}} \textbf{\bibinfo{volume}{572}},
  \bibinfo{pages}{211--214} (\bibinfo{year}{2019}).
%\newblock \eprint{1908.02372}.

\bibitem{Shu22}
\bibinfo{author}{{Shu}, X.} \emph{et~al.}
\newblock \bibinfo{title}{{A Census of Optically Dark Massive Galaxies in the
  Early Universe from Magnification by Lensing Galaxy Clusters}}.
\newblock \emph{\bibinfo{journal}{\apj}} \textbf{\bibinfo{volume}{926}},
  \bibinfo{pages}{155} (\bibinfo{year}{2022}).
%\newblock \eprint{2112.03709}.

\bibitem{Wang16}
\bibinfo{author}{{Wang}, T.} \emph{et~al.}
\newblock \bibinfo{title}{{Discovery of a Galaxy Cluster with a Violently
  Starbursting Core at $z = 2.506$}}.
\newblock \emph{\bibinfo{journal}{\apj}} \textbf{\bibinfo{volume}{828}},
  \bibinfo{pages}{56} (\bibinfo{year}{2016}).

\bibitem{McConachie21}
\bibinfo{author}{{McConachie}, I.} \emph{et~al.}
\newblock \bibinfo{title}{{Spectroscopic Confirmation of a Protocluster at
  $z=3.37$ with a High Fraction of Quiescent Galaxies}}.
\newblock %\emph{\bibinfo{journal}{arXiv}}
  \bibinfo{pages}{arXiv:2109.07696}
  (\bibinfo{year}{2021}).

\bibitem{Zavala19}
\bibinfo{author}{{Zavala}, J.~A.} \emph{et~al.}
\newblock \bibinfo{title}{{On the Gas Content, Star Formation Efficiency, and
  Environmental Quenching of Massive Galaxies in Protoclusters at z
  {\ensuremath{\approx}} 2.0-2.5}}.
\newblock \emph{\bibinfo{journal}{\apj}} \textbf{\bibinfo{volume}{887}},
  \bibinfo{pages}{183} (\bibinfo{year}{2019}).

\bibitem{Chapman09}
\bibinfo{author}{{Chapman}, S.~C.} \emph{et~al.}
\newblock \bibinfo{title}{{Do Submillimeter Galaxies Really Trace the Most
  Massive Dark-Matter Halos? Discovery of a High-z Cluster in a Highly Active
  Phase of Evolution}}.
\newblock \emph{\bibinfo{journal}{\apj}} \textbf{\bibinfo{volume}{691}},
  \bibinfo{pages}{560--568} (\bibinfo{year}{2009}).

\bibitem{Hung16}
\bibinfo{author}{{Hung}, C.-L.} \emph{et~al.}
\newblock \bibinfo{title}{{Large-scale Structure around a $z=2.1$ Cluster}}.
\newblock \emph{\bibinfo{journal}{\apj}} \textbf{\bibinfo{volume}{826}},
  \bibinfo{pages}{130} (\bibinfo{year}{2016}).

\bibitem{Shi19}
\bibinfo{author}{{Shi}, K.} \emph{et~al.}
\newblock \bibinfo{title}{{How Do Galaxies Trace a Large-scale Structure? A
  Case Study around a Massive Protocluster at $Z = 3.13$}}.
\newblock \emph{\bibinfo{journal}{\apj}} \textbf{\bibinfo{volume}{879}},
  \bibinfo{pages}{9} (\bibinfo{year}{2019}).

\bibitem{Nantais17}
\bibinfo{author}{{Nantais}, J.~B.} \emph{et~al.}
\newblock \bibinfo{title}{{Evidence for strong evolution in galaxy
  environmental quenching efficiency between z = 1.6 and z = 0.9}}.
\newblock \emph{\bibinfo{journal}{\mnras}} \textbf{\bibinfo{volume}{465}},
  \bibinfo{pages}{L104--L108} (\bibinfo{year}{2017}).
%\newblock \eprint{1610.08058}.

\bibitem{Weaver22}
\bibinfo{author}{{Weaver}, J.~R.} \emph{et~al.}
\newblock \bibinfo{title}{{COSMOS2020: A Panchromatic View of the Universe to z
  10 from Two Complementary Catalogs}}.
\newblock \emph{\bibinfo{journal}{\apjs}} \textbf{\bibinfo{volume}{258}},
  \bibinfo{pages}{11} (\bibinfo{year}{2022}).
%\newblock \eprint{2110.13923}.

\bibitem{Cai17}
\bibinfo{author}{{Cai}, Z.} \emph{et~al.}
\newblock \bibinfo{title}{{Mapping the Most Massive Overdensities through
  Hydrogen (MAMMOTH). II. Discovery of the Extremely Massive Overdensity
  BOSS1441 at $z = 2.32$}}.
\newblock \emph{\bibinfo{journal}{\apj}} \textbf{\bibinfo{volume}{839}},
  \bibinfo{pages}{131} (\bibinfo{year}{2017}).

\bibitem{Shi21}
\bibinfo{author}{{Shi}, D.~D.} \emph{et~al.}
\newblock \bibinfo{title}{{Spectroscopic Confirmation of Two Extremely Massive
  Protoclusters, BOSS1244 and BOSS1542, at $z = 2.24$}}.
\newblock \emph{\bibinfo{journal}{\apj}} \textbf{\bibinfo{volume}{915}},
  \bibinfo{pages}{32} (\bibinfo{year}{2021}).
\end{thebibliography}
\end{document}